\documentclass[10pt]{article}
\usepackage[dvips]{graphicx}
\usepackage[tight]{subfigure}
\usepackage{amssymb}
\usepackage{blindtext}
\usepackage{geometry}
\def\be{\begin{equation}}
\def\ee{\end{equation}}
\def\bea{\begin{eqnarray}}
\def\eea{\end{eqnarray}}
\def\beaN{\begin{eqnarray*}}
\def\eeaN{\end{eqnarray*}}
\def\ed{\end{document}}
\def\bit{\begin{itemize}}
\def\eit{\end{itemize}}

\def\sig{\sigma}

\def\k{\kappa}
\def\alf{\alpha}

\def\di{\partial}

\def\half{{\textstyle{1 \over 2}}}
\def\~{\tilde}
\def\lag{{{\cal L}}}
\def\m{\label}
\def\l{\left}
\def\r{\right}

\def\goto{\rightarrow}

\def\const{\rm const}
\def\diag{\rm diag}

\def\sA{{\stackrel{\bullet}{A}}{}}

\def\sR{\stackrel{\bullet}{R}}
\def\sG{\stackrel{\bullet}{\Gamma}}
\def\cG{\stackrel{\circ}{\Gamma}}

\def\sS{{\stackrel{\bullet}{S}}{}}

\def\sD{\stackrel{\bullet}{\cal D}}
\def\sJ{\stackrel{\bullet}{J}}

\def\sK{{\stackrel{\bullet}{K}}{}}
\def\sT{{\stackrel{\bullet}{T}}{}}

\def\sL{\stackrel{\bullet}{\lag}}
\def\cL{\stackrel{\circ}{\lag}}
\def\cN{\stackrel{\circ}{\nabla}}
\def\cA{{\stackrel{\circ}{A}}{}}
\def\cR{\stackrel{\circ}{R}}

\def\scJ{\stackrel{\bullet}{\cal J}}
\def\ccK{\stackrel{}{\cal K}}
\def\ccD{\stackrel{}{\cal D}}
\def\ccL{\stackrel{}{\cal L}}
\def\stheta{\stackrel{\bullet}{\theta}}

\usepackage{authblk}

\usepackage[dvipsnames]{xcolor}
\usepackage{graphicx}

\begin{document}

\title{ \bf  The equivalence principle for a plane gravitational wave in torsion based and non-metricity based teleparallel equivalents of general relativity}
\author[1]{E. D. Emtsova\thanks{Electronic address: \texttt{ed.emcova@physics.msu.ru}}}
\author[2]{A. N. Petrov\thanks{Electronic address: \texttt{alex.petrov55@gmail.com}}}
\author[1,2]{A. V. Toporensky\thanks{Electronic address: \texttt{atopor@rambler.ru}}}

\affil[1]{Kazan Federal University, Kremlevskaya 18, Kazan, 420008, Russia}
\affil[2]{Sternberg Astronomical institute, MV Lomonosov State university  \protect\\ Universitetskii pr., 13, Moscow, 119992,
Russia}

\date{\small \today}
\maketitle

\begin{abstract}

We study the energy-momentum characteristics of the plane ``+''-polarised gravitational wave solution of general relativity in the Teleparallel Equivalent of General Relativity (TEGR) and the Symmetric Teleparallel Equivalent of General Relativity (STEGR) using the previously constructed Noether currents. The current components describe locally measured by observer energy-momentum if the displacement vector $\xi$ is equal to the observer's 4-velocity. To determine the non-dynamical connection in these theories we use the unified ``turning off'' gravity principle. For a constructive analysis of the values of Noether currents and superpotentials in TEGR and STEGR, we use the concept of ``gauges''. The gauge changing can affect the Noether current values.  We study under what conditions the Noether current for the freely falling observer is zero. When they are established, zero result can be interpreted as a correspondence to the equivalence principle, and it is a novelty for gravitational waves in TEGR and STEGR.
We highlight two important cases with positive and zero energy, which reproduce the results of previous works with a different approach to determine gravitational energy-momentum in TEGR, and give their interpretation.
\end{abstract}

\section{Introduction}
\m{Introduction}

Teleparallel theories of gravity have been actively developed in recent years. One of important features of these theories is the use of teleparallel connections with zero Riemann curvature. A notion of a teleparallel connection (flat connection) unites such notions as inertial spin connection, Weitzenb\"ock connection, symmetric teleparallel affine connection, etc., which are more appropriate in any concrete case.
These theories include the Teleparallel Equivalent of General Relativity (TEGR), the Symmetric Teleparallel Equivalent of General Relativity (STEGR) and modifications of these theories \cite{BeltranJimenez:2019tjy,Heisenberg:2018vsk,Aldrovandi_Pereira_2013,BeltranJimenez:2019odq,Bahamonde:2021gfp,Adak:2023ymc}. In TEGR and its modifications, a flat metric compatible connection  with non-zero torsion is used. In STEGR and its modifications, a flat metric non-compatible connection with zero torsion is used. TEGR and STEGR are fully equivalent to  General Relativity (GR) at the level of field equations, thus, the solutions of the field equations in TEGR and STEGR are exactly the same as those in GR. In many cases, modifications of TEGR and STEGR  have the advantage that their field equations are of the second-order, what gives similarities with gauge field theories and potentially links gravity  to other theories of fundamental interactions in nature.

Because the field equations in TEGR and STEGR are equivalent to those in GR they do not include the teleparallel connections in whole. Formally it follows from the fact that the TEGR and STEGR Lagrangians differ from the Hilbert Lagrangian by divergences, only which include teleparallel connections, see \cite{Golovnev:2017dox} and formulae (\ref{lag+div}) and (\ref{Ls}) below. Thus, varying the TEGR and STEGR Lagrangians with respect to the teleparallel connections one obtains the identities $0=0$ which do not permit to determine them. In another words,
the teleparallel connections in TEGR and STEGR are not dynamical quantities and do not influence the dynamics of  gravitational interacting objects.

Teleparallel connections (as external fields) can be used in order to represent the TEGR equations in evidently covariant form with respect to Lorentz rotations of tetrads (where coordinate covariance already holds), or to represent the STEGR equations in evidently coordinate covariant form with respect to symmetric affine teleparallel connection. As usual, covariantized equations are used to construct conserved quantities, see, for example, the book \cite{Aldrovandi_Pereira_2013}. However, such a method has big difficulties which are described in detail in the framework of TEGR  in the conference presentation \cite{Krssak} by Kr\v s\v s\'ak where the results of previous studies of many authors are summarised. Let us list more important Kr\v s\v s\'ak's requirements for constructing energy-momentum in TEGR. It has to 1) be of the first derivatives only, 2) be covariant with respect to both coordinate transformations and local Lorentz rotations, 3) permit to construct global (integral) conserved quantities, or conserved charges. The most of known variants of energy-momentum in TEGR satisfy the first requirement. However, the requirements 2) and 3) are not consistent as a rule. On the one hand, one has well defined conserved charges expressed through well defined surface integrals, but one has only Lorentz non-covariant conserved energy-momentum. On the other hand, one can construct Lorentz covariant conserved energy-momentum, but then the related conserved charges cannot be defined.

In the framework of TEGR, in papers \cite{Obukhov:2006ge,Obukhov_Rubilar_Pereira_2006}, using Noether theorem, the aforementioned problem is resolved and fully covariant conserved quantities are constructed in the formalism of differential forms. However, this formalism is not so popular and, unfortunately, the papers \cite{Obukhov:2006ge,Obukhov_Rubilar_Pereira_2006} did not receive a relevant development.
 In \cite{EPT19,EPT_2020}, we have constructed fully covariant conserved quantities in TEGR (including well defined charges) in the tensorial formalism by direct application of the Noether theorem.
 The success was achieved thanks to including the teleparallel connection (inertial spin connection) as well as preserving displacement vectors in expressions after applying the Noether theorem to the diffeomorphically invariant TEGR action. Choosing displacement vectors as proper vector of observers, or Killing spacetime vectors, etc., one defines a character of conserved quantities. Our approach differs from others in that the observer is associated with a displacement vector rather than a timelike tetrad vector.

 Concerning STEGR, we do not feel an extensive study in construction of conserved quantities.  Nevertherless, in \cite{BeltranJimenez:2021kpj,Gomes:2022vrc}
 covariant Noether conserved quantities were constructed for general metric-affine gravity including TEGR and STEGR. However their correspondence to physically expected values was not studied. In \cite{EPT:2022uij}, following the ideology of \cite{EPT19,EPT_2020}, we have constructed covariant conserved quantities as well.

The new expressions for conserved quantities constructed in \cite{EPT19,EPT_2020,EPT:2022uij} have been used in various applications. It was shown that Noether's current is in correspondence with the weak equivalence principle for the freely falling observers ``frozen'' into the Hubble flow in Friedmann–Lemaître–Robertson–Walker universe and (anti-)de Sitter space, and Noether's charge gives correct mass for the Schwarzschild black hole.  In \cite{EKPT_2021,EKPT_2021a}, in TEGR, the new Noether's conserved quantities were studied in more detail and ambiguities of a special  character have been clarified as follows. These conserved quantities contain teleparallel connections evidently and they cannot be suppressed like in field equations in whole. But they are not dynamic variables of the theory and cannot be defined inside TEGR and STEGR themselves. Usually, authors use a so-called principle of ``turning off'' (``switching off'') gravity, see, for example, \cite{Obukhov+,Krssak:2015oua}. In \cite{EPT19,EPT_2020} we have generalized this principle in TEGR. In \cite{EPT:2022uij}, we have used it in STEGR, although it was not formulated there in fact, we formulate it here.

It turns out that even the generalized principle of ``turning off'' gravity cannot determine teleparallel connections by an unique way suppressing ambiguities. In TEGR, we study such a problem in more detail in   \cite{EKPT_2021,EKPT_2021a} on the example of the Schwarzschild solution. It was considered 1) a static tetrad with a related teleparallel connection and 2) a freely falling tetrad with another teleparallel connection. In the first case, static observers at spacelike infinity measure the standard mass of the black hole, and, in the second case, freely falling observers measure zero energetic characteristics that corresponds to the weak equivalence principle. Thus, in both the cases one has acceptable results. However, if one considers a freely falling observer together with the first pair of tetrad and connection, and if one considers a static distant observer together with the second pair of tetrad and connection,  one does not obtain acceptable results in each of the cases. It can be interpreted only that each of concrete tasks requires an appropriate pair of  tetrad and teleparallel connection.

M{\o}ller \cite{Moller_1961}, constructing covariant energy-momentum for gravitational field in the tetrad form instead of pseudotensors, has clarified that, being coordinate covariant, it is not covariant with respect to local Lorentz rotations. This problem has been resolved when teleparallel connection (inertial spin connection) is incorporated into consideration, however, the problems accented by  Kr\v s\v s\'ak  \cite{Krssak} appear. Such problems have been resolved as well \cite{Obukhov:2006ge,Obukhov_Rubilar_Pereira_2006,EPT19,EPT_2020,EPT:2022uij}, however, the problem of ambiguity in determination of teleparallel connection remains. How can it be studied? Each concrete solution and each concrete task require a separate approach. For example, in \cite{Hohmann:2019nat,Hohmann:2020,Bahamonde_2021,DAmbrosio:2021zpm} authors consider solutions with spherical, cylindrical, cosmological symmetries in modified theories (it can be easily adopted to TEGR and STEGR), and for each of symmetries they develop a concrete derivation. Such a situation is analogous to the consideration in the standard metric presentation of GR. For example, Katz, Bichak and Lynden-Bell  \cite{KBL_1997} suggested  a bi-metric representation of GR where conserved quantities become covariant ones instead of classical pseudotensors. A non-covariance of the latter is connected with non-localizability of energy, momentum, etc., in GR \cite{Misner_Thorn_Wheeler_1973}. However, covariant quantities in \cite{KBL_1997} essentially depend on a choice of a background metric that is another manifestation of the non-localizability. As a result, a background metric has to be chosen for concrete solutions separately. A detailed comparison of the approach in \cite{KBL_1997} with constructions in TEGR, where a role of the background metric is played by teleparallel connections, is given in \cite{EPT19}. Thus, a reasonable (possible) interpretation of appeared problems with a choice of teleparallel connections in TEGR and STEGR can be connected with the non-localizability of conserved quantities in GR and cannot be avoided in principle.

In literature, various approaches to construct conserved quantities both in TEGR and in STEGR and their modifications have already been tested for the Schwarzschild solution and cosmological models, see, for example, \cite{Maluf_2005,Maluf0704,Obukhov+,Obukhov_2006,Obukhov_Rubilar_Pereira_2006,Bahamonde:2022zgj,BeltranJimenez:2019bnx,Gomes:2022vrc,Gomes:2023hyk} and references therein.
However, for the best of our knowledge, construction of conserved quantities for the gravitational waves in teleparallel gravity did not attract much attention of researches. We could cite the central papers with such study in TEGR only \cite{Maluf:2003fg,Maluf:2008yy,Obukhov:2009gv,Formiga:2018maj,Formiga_2020} with a few references therein. In calculations the energy-momentum tensor of gravitational field \cite{Aldrovandi_Pereira_2013} is used there.  By these studies,  authors sometimes got  unexpected results. Thus, for example, in \cite{Maluf:2008yy} non-positive energy for gravitational waves is obtained; on the other side, in  \cite{Obukhov:2009gv} zero energy for gravitational waves is obtained.
Such results are not acceptable; indeed, the modern cosmological and astrophysical observable data supports the textbook predictions that gravitational waves bring a positive energy \cite{Manchester:2015mda,  Antoniadis:2014eia, Esposito-Farese:2004pem}.
At last, we have not found in a literature any works where the problem of constructing conserved quantities for gravitational waves have been considered in STEGR.

To define energy and momentum densities of gravitational wave (or charges for it) one has to introduce observers, whose proper vectors are chosen to be corresponding displacement vectors. However, it is not so clear how one can provide this study in the formalism under consideration. Indeed, the gravitational wave spacetime has no a timelike Killing vector, and one cannot construct charges for static distant observers because it is not clear how they have to be defined. Nevertheless, in subsection 3.4, we discuss the problem of constructing energy and momentum densities of gravitational wave in the framework of our fully covariant formalism and give a related interpretation that differs from the interpretation in \cite{Formiga:2018maj, Formiga_2020}.

The importance to define a correspondence to the equivalence principle for various solutions was remarked many times, see, for example, the recent paper \cite{Vitorio_2022} and references therein.
For black holes and cosmological models such a correspondence has been stated, for example, in \cite{9,Maluf0704,EPT19,EKPT_2021,EPT:2022uij}. However, up to now it was not stated a correspondence to equivalence principle for gravitational wave solutions (although such attempts have been provided \cite{Formiga:2018maj, Formiga_2020}).

Unlike determining a timelike Killing vector for the gravitational wave, or determining distant observer for gravitational wave filling infinite space, it is easy to take a displacement vector as a freely falling observer's proper vector.  Then it remains only to determine the teleparallel connections that give the Noether current corresponding to the equivalence principle. Such a task can be resolved by correspondent efforts. Thus,  the more important purpose  of this paper is to achieve a correspondence with the equivalence principle. We consider only plane exact gravitational wave in TEGR and STEGR with only the one ``+'' polarization and with making the use of the fully covariant formalism developed by us \cite{EPT19,EKPT_2021,EPT:2022uij}.

The paper is organized as follows:

In section 2, we give brief introduction into TEGR and STEGR theories, introduce the Noether currents and superpotentials for these theories and describe the unified ``turning off'' gravity principle to define the teleparallel connection.

In section 3, we calculate energy and momentum for the plane ``+'' polarised gravitational wave measured by the freely falling observers in TEGR and STEGR in the simplest frames. In this case, we  obtain the result that does not correspond to equivalence principle. In the framework of the formalism \cite{EPT19,EKPT_2021,EPT:2022uij} such a result can be explained as the found teleparallel connections do not correspond to the equivalence principle for the chosen freely falling observers. However, at least in linear approximation, the result coincides with the Landau-Lifshitz formula for the energy density and energy density flux. This coincidence is interpreted.

In section 4, we set a task to find teleparallel connections, substitution of which into the Noether current expression for freely falling observers in TEGR and STEGR gives zero. This goal has been achieved and, thus, a correspondence with the equivalence principle has been stated.

In section 5, using the results of \cite{Obukhov:2009gv}, we take other coordinates (not simplest ones) for the plane ``+'' polarised gravitational wave, after that the equivalence principle for freely falling observers is established by another way.

In section 6, we discuss the results.

In Appendix A related to subsection 4.1, we state that for an arbitrary Lorentz rotation depending only on retarded time there are no compositions of such rotations of the tetrad only preserving inertial spin connection (or of inertial spin connection only preserving tetrad), which can change the Noether current.

In Appendix B related to section 5, we show that the tetrad considered in \cite{Obukhov:2009gv} is not a freely falling tetrad and, thus, zero result in \cite{Obukhov:2009gv} is not in correspondence with the equivalence principle.

All definitions in TEGR correspond to \cite{Aldrovandi_Pereira_2013},
all definitions in STEGR correspond to \cite{BeltranJimenez:2019tjy}.

\section{Preliminaries}
\m{Preliminaries}

In this section, we briefly introduce the elements of teleparallel equivalents of GR (TEGR and STEGR) and, following our papers \cite{EPT_2020,EKPT_2021,EPT:2022uij} give totally covariant expressions for conserved quantities, which are necessary for our calculations.

\subsection{Elements of TEGR}
\m{ElementsT}

The Lagrangian in TEGR has a form \cite{Aldrovandi_Pereira_2013}
\be
\sL =   \frac{h}{{2}\kappa}\sT\,
\equiv
\frac{h}{2\kappa} \l(\frac{1}{4} {\sT}{}^\rho{}_{\mu\nu} {\sT}_\rho{}^{\mu\nu} + \frac{1}{2} {\sT}{}^\rho{}_{\mu\nu} {\sT}{}^{\nu\mu}{}_\rho - {\sT}{}^\rho{}_{\mu\rho} {\sT}{}^{\nu\mu}{}_\nu\r),
\m{lag}
\ee
where $\sT$ is called as the torsion scalar, $\kappa=8 \pi$ in $c=G=1$ units.  Since the main goal of the present paper is to consider gravitational waves in  vacuum only, we do not add to (\ref{lag}) a matter Lagrangian. The torsion tensor ${\sT}{}^a{}_{\mu\nu}$ is defined as
\begin{equation}\label{tor}
{\sT}{}^a{}_{\mu\nu} = \di_\mu h^a{}_\nu - \di_\nu h^a{}_\mu + {\sA}{}^a{}_{c\mu}h^c{}_\nu - {\sA}{}^a{}_{c\nu}h^c{}_\mu,
\end{equation}
where $h^a{}_\nu$ are the tetrad components
 connected to metric by
\begin{equation}
    g_{\mu\nu}=\eta_{ab} h^a{}_\mu h^b{}_\nu\,
\m{g_munu}
\end{equation}
and $h = \det h^a{}_\nu$,
 we denote tetrad indexes of a quantity  by Latin letters   and  spacetime indexes by Greek letters.
It is useful to recall that the transformation of tetrad indices into spacetime ones and vice versa is performed by contraction with tetrad vectors, for example,  $\sT{}^\alf{}_{\mu\nu} =h^\alf{}_a \sT{}^a{}_{\mu\nu}$, etc.

From now and below, we denote by $\bullet$ all teleparallel quantities, which are constructed using the teleparallel affine connection (Weitzenb\"ock connection) $\sG{}{}^\alpha {}_{\kappa \lambda}$. Thus,
the inertial spin connection (ISC) ${\sA}{}^a{}_{c\nu}$ is defined as
\begin{equation}\label{ISCdef}
    \sA{}^a{}_{b\mu} = -h_b{}^\nu \stackrel{\bullet}{\nabla}_\mu h^a{}_\nu,
\end{equation}
where the covariant derivative $\stackrel{\bullet}{\nabla}_\mu$ corresponds to $\sG{}{}^\alpha {}_{\kappa \lambda}$.
The connection $\sG{}{}^\alpha {}_{\kappa \lambda}$ is flat, i.e., the related curvature is equal to zero:
\begin{equation}\label{RiemTEGR}
    \sR{} {}^\alpha {}_{\beta \mu \nu} \equiv \di_\mu \sG{}^\alpha {}_{\beta\nu} - \di_\nu \sG{}^\alpha{}_{\beta \mu} + \sG{}^\alpha{}_{\kappa \mu}\sG{}^\kappa {}_{\beta \nu} - \sG{}^\alpha{}_{\kappa \nu}\sG{}^\kappa {}_{\beta \mu}=0\,.
\end{equation}
The connection $\sG{}{}^\alpha {}_{\kappa \lambda}$ is also compatible with the physical metric, that is, the corresponding non-metricity is zero:
\begin{equation}
 \stackrel{\bullet}{Q}   {}_{\mu \alf \beta}  \equiv  \stackrel{\bullet}{\nabla}{}_\mu g_{\alf \beta}=0.
 \m{bulletQ}
\end{equation}

In our notations, following to \cite{Aldrovandi_Pereira_2013}, quantities denoted by a $\circ$ are constructed with the use of the Levi-Civita affine connection $\cG{}{}^\alpha {}_{\kappa \lambda}$. Thus,
$\cA^a{}_{b\rho} $ is the usual  Levi-Civita spin connection (L-CSC) defined by
\begin{equation}\label{A}
    \cA{}^a{}_{b\mu} = -h_b{}^\nu \cN_\mu h^a{}_\nu,
\end{equation}
where the covariant derivative $ \cN_\mu$ is constructed with  $\cG{}{}^\alpha {}_{\kappa \lambda}$. Now it is useful to introduce a contortion tensor defined as:
  \be
\sK^a{}_{b\rho} = \sA^a{}_{b\rho} - \cA^a{}_{b\rho}.
\m{K_A_A}
\ee

Let us discuss the role of the ISC incorporated into the torsion tensor (\ref{tor}). From the beginning let us rewrite the Lagrangian (\ref{lag}) in the other form \cite{Aldrovandi_Pereira_2013}:
\be
{\sL} =   {\cL} -\frac{1}{\kappa}\di_\mu\l(h{\sT}{}^{\nu\mu}{}_\nu\r)\,,
\m{lag+div}
\ee
with the Hilbert Lagrangian
\be
{\cL} =  -\frac{h}{2\kappa}\cR\,,
\m{lag_H}
\ee
where $\cR$ is the Riemannian curvature scalar presented as a function of tetrad components in correspondence with (\ref{g_munu}), see \cite{Landau_Lifshitz_1975}. Thus,  the TEGR Lagrangian contains ISC in the divergence only. Then, first, varying the action with the Lagrangian (\ref{lag}), the same (\ref{lag+div}), with respect to tetrad components one obtains the Euler-Lagrange equation
 \be
E_a{}^\rho \equiv \frac{\delta \sL}{\delta h^a{}_\rho} \equiv \frac{\di \sL}{\di h^a{}_\rho} - \di_\sig \l(\frac{\di \sL}{\di h^a{}_{\rho,\sigma}} \r)=0,
 \m{EM+}
 \ee
 where $E_a{}^\rho$ does not actually depend on ISC because
 \be
E_a{}^\rho \equiv \frac{\delta \sL}{\delta h^a{}_\rho} \equiv \frac{\delta \cL}{\delta h^a{}_\rho}=0\,.
 \m{EMc}
 \ee
 This means also that TEGR and GR are equivalent.
 Second, varying the action with the Lagrangian (\ref{lag}), the same (\ref{lag+div}), with respect to $\sA^a{}_{b\rho} $ (that is included in the divergence only) one obtains $0=0$. Thus, ISC cannot be determined in the framework of TEGR itself. Then, being an external structure, it can be defined by additional requirements only.

 However, what is the role of $\sA^a{}_{b\rho} $? At least its presence allows us to present the torsion tensor in a covariant form with respect to local Lorentz rotations, see (\ref{tor}). This form gives a possibility to represent the following tensors in fully and evidently covariant form. The contortion tensor can be rewritten in the convenient form:
    \begin{equation}\label{tor_K}
    \sK^\rho{}_{\mu\nu}=\frac{1}{2}(\sT_\mu{}^\rho{}_\nu+\sT_\nu{}^\rho{}_\mu -\sT^\rho{}_{\mu\nu}).
\end{equation}
 The torsion scalar  in (\ref{lag}) can be rewritten in the form:
\begin{equation}\label{torscalar}
\sT=\frac{1}{2}\,{\sS}_a{}^{\rho\sigma}{\sT}{}^a{}_{\rho\sigma},
\end{equation}
where the teleparallel superpotential ${\sS}_a{}^{\rho\sigma}$  defined as
\begin{equation}\label{super_K}
     {\sS}_a{}^{\rho\sigma}=
     %- \frac{\kappa}{h} \frac{\di \sL}{\di h^a{}_{\rho,\sigma}}
     \sK{}^{\rho\sigma} {}_a + h_a{}^{\sigma} \sK{}^{\theta \rho} {}_{\theta} - h_a{}^{\rho} \sK{}^{\theta \sigma} {}_{\theta}\,
\end{equation}
 is an antisymmetric tensor in the last two indices.

All tensors ${\sT}{}^a{}_{\mu\nu}$, $\sK^{\rho\sigma}{}_{a}$, and ${\sS}_a{}^{\rho\sigma}$, are covariant with respect to both coordinate transformations and local Lorentz transformations. (For tensors without Lorentz indexes or scalars we say ``invariant  with respect to  Lorentz transformations''.) Note, that working in the covariant formulation of teleparallel gravity, the local Lorentz covariance means that the tensorial quantities are transformed covariantly under the simultaneous transformation of both the tetrad and the ISC:
\begin{equation}\label{lroth}
h'^a {}_{\mu} = \Lambda^a {}_b (x) h^b {}_\mu\,,
\end{equation}
\begin{equation}\label{spin_trans}
\sA{}'{}^a {}_{b \mu}=\Lambda {}^a {}_c  (x) \sA{} {}^c {}_{d \mu} \Lambda {}_b {}^d   (x)  + \Lambda {}^a {}_c  (x) \partial_\mu \Lambda {}_b {}^c  (x) ,
\end{equation}
where $\Lambda {}^a {}_c  (x)$ is the matrix of a local Lorentz rotation, and $\Lambda {}_a {}^c  (x)$  is an inverse matrix  of the latter. The operation at the right hand side of (\ref{spin_trans}) tells us that ISC can be equalized to zero by an appropriate local Lorentz transformation. Then, by another local Lorentz rotation it can be represented in the form:
\begin{equation}\label{telcon}
 {\sA}{}^a{}_{c\nu}=\Lambda^a {}_b \partial_\nu    (\Lambda^{-1}){}^b {}_c.
\end{equation}

 Let us return to the field equations (\ref{EM+}). With zero matter part, they can be rewritten as
\begin{equation}\label{PereiraCurrdivsup}
 \kappa  h\sJ_{a}{}^{\rho}   =\partial_\sigma\Big(h\sS_a{}^{\rho \sigma}\Big) = \cN_\sigma\Big(h\sS_a{}^{\rho \sigma}\Big),
\end{equation}
where the gravitational energy-momentum $\sJ_{a}{}^{\rho}$ is defined as
  \begin{equation}\label{PereiraCurr}
\sJ_{a}{}^{\rho}= \frac{1}{\k}h_a{}^\mu \sS_c{}^{\nu\rho}\sT^c{}_{\nu\mu} -\frac{h_a{}^\rho}{h}\sL + \frac{1}{\k}\sA^c{}_{a\sig}\sS_c{}^{\rho\sig}\,,
\end{equation}
in \cite{Aldrovandi_Pereira_2013} it is called as a gravitational current as well. The teleparallel superpotential (\ref{super_K}) is antisymmetric in upper indexes, therefore the current density of (\ref{PereiraCurr}) is conserved $\di_\rho(h\sJ_{a}{}^{\rho})=0$. Because both the l.h.s. and the r.h.s. of (\ref{PereiraCurrdivsup}) are coordinate covariant, one can define a well defined charges. However, the current $\sJ_{a}{}^{\rho}$ itself is not Lorentz covariant that is a problem itself.

Because (\ref{PereiraCurrdivsup}) in whole is Lorentz covariant one has a possibility to construct  both l.h.s. and r.h.s. in evidently Lorentz covariant form as well. Thus, (\ref{PereiraCurrdivsup}) can be rewritten as
\begin{equation}\label{PereiraCurrdivsup+}
 \kappa  h{\sJ}{}'_{a}{}^{\rho}   =\sD_\sigma\Big(h\sS_a{}^{\rho \sigma}\Big),
\end{equation}
where the Lorentz covariant current is defined as \cite{Aldrovandi_Pereira_2013}
  \begin{equation}\label{PereiraCurr+}
{\sJ}{}'_{a}{}^{\rho}= \frac{1}{\k}h_a{}^\mu \sS_c{}^{\nu\rho}\sT^c{}_{\nu\mu} -\frac{h_a{}^\rho}{h}\sL\,,
\end{equation}
and $\sD_\sigma$ is the Lorentz covariant derivative. For example, applied to a tetrad co-vector, $V_a$, it is defined as
\be
\sD_\sigma V_a = \di_\sigma V_a - \sA{}^c{}_{a\sigma} V_c\,.
\m{sD}
\ee
However, the form (\ref{PereiraCurrdivsup+}) does not allows us to construct well defined charges. It is just the problem accented by Kr\v s\v s\'ak \cite{Krssak} that has been resolved by introducing our formalism \cite{EPT19,EPT_2020}.

How it was noted the conserved quantities suggested in \cite{EPT19,EPT_2020} are obtained by applying the Noether theorem that is they can be classified as canonical conserved quantities. This means that the divergences in Lagrangians have to be taken into account. Thus, considering the form of the TEGR Lagrangian (\ref{lag+div}), one concludes that the conserved quantities contain the ISC in an explicit form, so their values are sensitive to the ISC changing. Thus, some external procedure is needed to restrict ISC so that the conserved quantities  would have physically meaningful values, see in more detail subsections \ref{CL} and \ref{turning_off}.

\subsection{Elements of STEGR}
\m{ElementsS}

The Lagrangian in STEGR has the form \cite{BeltranJimenez:2019tjy}:
\begin{equation}\label{STlag}
  \ccL{} = \frac{\sqrt{-g}}{2 \kappa} g^{\mu\nu} (L^{\alpha} {}_{\beta \mu} L^{\beta} {}_{\nu \alpha} - L^{\alpha} {}_{\beta \alpha} L^{\beta} {}_{\mu\nu} ),
\end{equation}
where again $\kappa = 8\pi$ and $g=\det g_{\mu\nu}$.  The disformation tensor $ L^{\alpha} {}_{\mu \nu}$ is defined as:
\begin{equation}\label{defLofQ}
    L^{\alpha} {}_{\mu \nu}=\frac{1}{2} Q^{\alpha} {}_{\mu \nu} -\frac{1}{2} Q_{\mu} {}^{\alpha} {}_{\nu}-\frac{1}{2} Q_{\nu} {}^{\alpha} {}_{\mu}\,,
\end{equation}
and the non-metricity tensor $Q_{\alpha \mu \nu}$ is defined as follows:
\begin{equation}\label{defQ}
    Q_{\alpha \mu \nu} \equiv \nabla_\alpha g_{\mu \nu}
\end{equation}
that is not zero in general, unlike (\ref{bulletQ}).
Here, the covariant derivative $\nabla_\alpha$ is defined with the use of the flat affine connection $\Gamma^{\alpha} {}_{\mu \nu}$ which is symmetric in lower indexes, we denote it below by abbreviation STC --- symmetric teleparallel connection. The corresponding torsion is zero: $T {}^\alpha {}_{\mu \nu} \equiv \Gamma {}^\alpha {}_{\mu \nu} - \Gamma {}^\alpha {}_{\nu\mu } = 0$. The curvature tensor for STC is zero as well:
\begin{equation}\label{defRiem}
    R^{\alpha} {}_{\beta \mu \nu} (\Gamma) = \partial_{\mu} \Gamma^{\alpha} {}_{ \nu \beta} -  \partial_{\nu} \Gamma^{\alpha} {}_{\mu \beta} +  \Gamma^{\alpha} {}_{\mu \lambda}  \Gamma^{\lambda} {}_{\nu \beta} -  \Gamma^{\alpha} {}_{\nu \lambda}  \Gamma^{\lambda} {}_{\mu \beta} =0.
\end{equation}
Thus, in the framework of STEGR, the metric components are thought as dynamic variables and gravitational effects are encoded in the non-metricity (\ref{defQ}).

One can easily verify that the decomposition of a general connection $\Gamma^{\beta} {}_{\mu \nu}$ into the Levi-Civita connection, contortion and the disformation terms, see \cite{BeltranJimenez:2019tjy}, reduces to
\begin{equation}\label{defL}
    L^{\beta} {}_{\mu \nu} \equiv \Gamma^{\beta} {}_{\mu \nu} - \cG{}^{\beta} {}_{\mu \nu}.
\end{equation}
Here, the Levi-Civita connection $\cG{}^{\beta} {}_{\mu \nu}$ is considered as a function of metric in the usual way \cite{Landau_Lifshitz_1975}. Then, with the use of (\ref{defLofQ}) - (\ref{defL}) one can rewrite (\ref{STlag}) as
\begin{equation}\label{STlagtoHilbert}
      \ccL{} =   \cL+ \frac{\sqrt{-g} g^{\mu\nu}}{2 \kappa} R_{\mu\nu} +   \ccL{}' .
\end{equation}
The first term is the Hilbert Lagrangian (\ref{lag_H}), however, presented here as a function of metric
\be
{\cL} =  -\frac{\sqrt{-g}}{2\kappa}\cR\,.
\m{lag_H+}
\ee
According to \cite{EPT:2022uij} we  neglect the second term in (\ref{STlagtoHilbert})\footnote{The second term in (\ref{STlagtoHilbert} is equal to zero due to (\ref{defRiem}). However, if we preserve it in the Lagrangian, we need to vary it. Thus, variation of (\ref{STlagtoHilbert}) together with the second term means the variation of (\ref{STlag}) exactly. But, variation of (\ref{STlag}) with respect to $\Gamma^{\beta} {}_{\mu \nu}$ gives $\Gamma^{\beta} {}_{\mu \nu} = \cG{}^{\beta} {}_{\mu \nu}$. This means that the flat STC $\Gamma^{\beta} {}_{\mu \nu}$ can be equal to the Levi-Civita connection that is not flat on general. Since it is not permissible the second term in (\ref{STlagtoHilbert}) has to be cancelled.}. The third term $\ccL'$ is a total divergence:
\begin{equation}\label{defD}
      \ccL'{} =   - \frac{\sqrt{-g}}{2 \kappa}  \cN_\alpha (Q^\alpha-\hat{Q}^\alpha)=   \partial_\alpha \l(- \frac{\sqrt{-g}}{2 \kappa}  (Q^\alpha-\hat{Q}^\alpha)\r)=\partial_\alf  \ccD{}^\alf,
\end{equation}
where $Q_\alpha=g^{\mu\nu} Q_{\alpha \mu \nu}$ and $\hat{Q}_\alpha=g^{\mu\nu} Q_{\mu \alpha \nu}$. Using (\ref{defLofQ}), one easily finds that
\begin{equation}\label{Q_L}
   Q^\alpha=-2 g^{\mu\nu} L_\mu{}^\alpha{}_\nu, \qquad \hat{Q}^\alpha=-g^{\mu\nu}\l(L^\alf{}_{\mu\nu} + L_\mu{}^\alpha{}_\nu\r)\,,
\end{equation}
where $L^\alf{}_{\mu\nu}$ is thought as defined in (\ref{defL}).
Thus, STC is included in divergence only, and we consider the Lagrangian
\begin{equation}\label{Ls}
    \ccL{} =  - \frac{\sqrt{-g}}{2 \kappa} \cR  + \partial_\alf  \ccD{}^\alf,
\end{equation}
 with
\begin{equation}\label{defD2}
    \ccD^{\alpha} \equiv  - \frac{\sqrt{-g}}{2 \kappa}  (Q^\alpha-\hat{Q}^\alpha)\,
\end{equation}
instead of (\ref{STlag}) or (\ref{STlagtoHilbert}).

Varying action with the Lagrangian (\ref{Ls}) with respect to $g^{\alf\beta}$ one obtains the vacuum field equations
\begin{equation}\label{Eqs_Q}
    \cR_{\alf\beta}\l(\delta^\alf_\mu\delta^\beta_\nu -\half g_{\mu\nu}g^{\alf\beta} \r) = 0,
\end{equation}
where
\begin{equation}\label{Rmunu_Q}
    \cR_{\alf\beta} = \di_\rho\cG{}^\rho{}_{\alf\beta} - \di_\alf\cG{}^\rho{}_{\rho\beta} + \cG{}^\rho{}_{\alf\beta}\cG{}^\sig{}_{\sig\rho} - \cG{}^\rho{}_{\alf\sig}\cG{}^\sig{}_{\beta\rho}\,.
\end{equation}
Because the equations (\ref{Eqs_Q}) are the usual GR equations (which do not contain STC in whole) we are convinced that GR and STEGR are equivalent. The reason of absence of STC in (\ref{Eqs_Q}) consists in the fact that STC is included in the divergence in (\ref{Ls}) only. However, the expression for the Ricci tensor (\ref{Rmunu_Q}), containing partial derivatives, is not evidently covariant, at least, one has to prove it additionally. Thus, the same expression (\ref{Rmunu_Q}) can be  rewritten with the use of STC in evidently covariant form:
\begin{equation}\label{Rmunu_QL}
    \cR_{\alf\beta} = \nabla_\alf L^\rho{}_{\rho\beta} -  \nabla_\rho L^\rho{}_{\alf\beta}+ L^\rho{}_{\alf\beta} L^\sig{}_{\sig\rho} - L^\rho{}_{\alf\sig}L^\sig{}_{\beta\rho}\,,
\end{equation}
where $L^{\alf} {}_{\mu \nu}$ defined in (\ref{defL}) is  a tensor. The situation is analogous to the one in TEGR, where the field equations (\ref{PereiraCurr+}) are evidently and fully covariant after incorporation of ISC into them.

Let us repeat, the STEGR Lagrangian contains STC in the divergence only. By this, first, varying the action with the Lagrangian (\ref{Ls}) with respect to the metric components one obtains the Euler-Lagrange equations (\ref{Eqs_Q}). Second, varying the action with the Lagrangian (\ref{Ls}) with respect to $\cG{}^\alf{}_{\mu\nu}$ (that is included in the divergence only) one obtains $0=0$. Summarizing, STC cannot be determined in the framework of STEGR itself. This means that STC, the same as ISC in TEGR, being an external structure, can be defined by additional requirements only.

On the other hand, the conserved quantities suggested in \cite{EPT:2022uij} are canonical ones because they are obtained by applying the Noether theorem. This means that the divergence in the Lagrangian (\ref{Ls}) has to be taken into account. As a result, conserved quantities contain STC in evident from that has to be determined by additional procedure, see in more detail subsections \ref{CL} and \ref{turning_off}.

\subsection{Noether conserved quantities}
\m{CL}

The Noether conserved quantities were derived for TEGR Lagrangian
 (\ref{lag}) in our papers \cite{EPT19,EPT_2020} and for STEGR Lagrangian  (\ref{Ls}) in another our paper \cite{EPT:2022uij}. In both theories Noether current ${\cal I}{}^{\alf}(\xi)$ is a vector density of the weight +1, and the Noether superpotential ${\cal I}{}^{\alf\beta}(\xi)$ is an antisymmetric tensor density  of the weight +1. For such quantities $\cN_\mu \equiv \partial_\mu$, and, thus, the conservation laws have evidently covariant  form:
  \be
\di_\alf {\cal I}{}^\alf(\xi) \equiv \cN_\alf {\cal I}{}^\alf(\xi) = 0\,.
\m{CL_current}
\ee
  \be
{\cal I}{}^\alf(\xi) = \di_\alf {\cal I}{}^{\alf\beta}(\xi) \equiv \cN_\alf {\cal I}{}^{\alf\beta}(\xi)\,.
\m{CL_c_s}
\ee
It is important to give a physical interpretation of the quantities presented above. We follow the prescription in \cite{EPT19,EPT:2022uij}. First, one has to choose the displacement vector $\xi^\alpha$, it can be Killing vector, proper vector of observer, etc. Second, setting a time coordinate as $t=x^0$ and choosing a space section as $\Sigma := t=\const$,
one can interpret ${\cal I}^0(\xi)$ as a density on the section $\Sigma$ of the quantity related to a chosen  $\xi^\alpha$. For example, if $\xi^\alpha$ is a time-like Killing vector it is interpreted as the energy density on $\Sigma$. On the other hand, if $\xi^\alpha$ is an observer's proper vector, then the components ${\cal I}^\alpha(\xi)$ can be interpreted as components of the energy-momentum vector measured by such an observer.

The Noether current ${\scJ}{}^{\alf}(\xi) $ in TEGR \cite{EPT19,EPT_2020} in vacuum case and when the field equations (\ref{EM+}) hold
acquires the form:
\begin{equation}\label{noethercurrtegr}
    {\scJ}{}^\alf(\xi) =  h\stheta_\sig{}^\alf
\xi^\sigma + \frac{h}{\k} \sS_{\sig}{}^{\alf\rho}\cN_\rho\xi^\sig\,,
\end{equation}
where the gravitational Noether energy-momentum tensor $\stheta_\sig{}^\alf$ is
\begin{equation}\label{theta}
     \stheta_\sig{}^\alf  \equiv \frac{1}{\k}\sS_{a}{}^{\alf\rho}{\sK}^a{}_{\sig\rho}- \frac{1}{h}\sL \delta^\alf_\sig \,.
\end{equation}
The related Noether superpotential in TEGR is
  \be
{\scJ}{}^{\alf\beta}(\xi) = \frac{h}{\k}{\sS}_a{}^{\alf\beta}h^a{}_\sigma\xi^\sig\,.
\m{super}
\ee
In correspondence with (\ref{CL_c_s}) one can derive the current from the superpotential:
\begin{equation}\label{curreqdivsupTEGR}
    {\scJ}{}^{\alf}(\xi) = \partial_\beta {\scJ}{}^{\alf\beta}(\xi).
\end{equation}
Both ${\sJ}{}^{\alf}(\xi)$ and ${\sJ}{}^{\alf\beta}(\xi)$ essentially depend on tensors ${\sT}{}^a{}_{\mu\nu}$, $\sK^{\rho\sigma}{}_{a}$ and ${\sS}_a{}^{\rho\sigma}$ defined in (\ref{tor}), (\ref{tor_K}) and (\ref{super_K}) and which are covariant with respect to both coordinate transformations and local Lorentz rotations. Thus, both  ${\sJ}{}^{\alf}(\xi)$ and ${\sJ}{}^{\alf\beta}(\xi)$ are explicitly spacetime covariant and Lorentz invariant.

Such an advantage is achieved  by the fact that ${\sT}{}^a{}_{\mu\nu}$, $\sK^{\rho\sigma}{}_{a}$ and ${\sS}_a{}^{\rho\sigma}$ contain the components of ISC which are transformed
simultaneously with the tetrad components, see (\ref{lroth}) and (\ref{spin_trans}). This means that if the pair of  tetrad and ISC is fixed one has  ${\sJ}{}^{\alf}(\xi)$ and ${\sJ}{}^{\alf\beta}(\xi)$ defined respectively to this pair. In \cite{EKPT_2021,EKPT_2021a}, keeping in mind this situation we introduce a notion of ``gauges'' in TEGR. A gauge is defined as a set of pairs  (tetrad and ISC) which can be obtained from a given combination of the tetrad and the ISC by an arbitrary Lorentz rotation where tetrad transforms as (\ref{lroth}) and  inertial spin connection transforms as (\ref{spin_trans}) simultaneously, and/or arbitrary coordinate transformations.
The case in which zero ISC corresponds to some tetrad is called traditionally as the Wietzenb\"ock gauge \cite{Aldrovandi_Pereira_2013}. In the above cases, the word ``gauge'' is used by us in different senses: when we say ``Wietzenb\"ock gauge'' we mean the only one  pair - tetrad and zero ISC; and when we say ``gauge'' in our definition we mean the whole equivalence class of pairs of tetrads and ISCs in which two pairs are equivalent if and only if they are connected as defined above.

In STEGR with the Lagrangian (\ref{Ls}) the conserved quantities were derived in \cite{EPT:2022uij} for each term separately.
The superpotential ${\cal J}{}_{GR}^{\alpha\beta}$ for the Hilbert term (\ref{lag_H}) is the Komar superpotential  \cite{Mitskevich_1969, Petrov_KLT_2017}
\begin{equation}\label{Komarsup}
     {\cal J}{}_{GR}^{\alpha\beta} =  \ccK{}^{\alpha\beta} = \frac{\sqrt{-g}}{\kappa} \cN{}^{[\alpha} \xi^{\beta]},
\end{equation}
For the divergent term $\ccL'{}= \partial_\alf  \ccD{}^\alf$, see (\ref{defD}) and (\ref{defD2}), the Noether superpotential is:
\begin{equation}\label{addsupstegr}
 {\cal J}{}_{div}^{\alpha\beta}= \frac{\sqrt{-g}}{\kappa} \delta_{\sigma}^{[\alpha}  (Q^{\beta]}-\hat{Q}^{\beta]}) \xi^\sigma.
\end{equation}
The total Noether  superpotential of the Lagrangian (\ref{Ls}) in STEGR is:
\begin{equation}\label{totalsup}
    {\cal J}{}^{\alpha\beta} = {\cal J}{}_{GR}^{\alpha\beta} + {\cal J}{}_{div}^{\alpha\beta}.
\end{equation}
Taking the divergence of each term of (\ref{totalsup}) in correspondence with (\ref{CL_c_s}) one gets the total Noether current
\begin{equation}\label{totalcurr}
    {\cal J}{}^{\alpha} = {\cal J}{}_{GR}^{\alpha} + {\cal J}{}_{div}^{\alpha}.
\end{equation}
As one can see $ {\cal J}{}^{\alf}(\xi)$ and $ {\cal J}{}^{\alf\beta}(\xi)$ are explicitly spacetime covariant.

In \cite{EPT:2022uij}, the concept of gauges in STEGR has not been introduced.
%, because there is no additional structure, such as a tetrad in TEGR. In fact, in STEGR, if we find different teleparallel connections $\Gamma {}^\alpha {}_{\mu\nu}$ in the same coordinates, this will lead to different values of the conserved quantities.
In the present paper we  want to have the same terminology in STEGR as in TEGR. So, formally, we take some specific pair of coordinates $\bar{x}^{\mu}$ and STC  $\bar\Gamma {}^\alpha {}_{\mu\nu}$, and then associate with this pair a class of pairs (${x}^{\mu}$,  $\Gamma {}^\alpha {}_{\mu\nu}$) which are  connected to it by the transformations
\bea\label{affconntransf}
{x}^{\mu} &=&  {x}^{\mu}(\bar{x}^\alf),\nonumber\\
  \Gamma {}^\alpha {}_{\mu\nu} &=&\frac{\partial x^\alpha}{\partial \bar{x}{}^{\bar{\alpha}}} \frac{\partial \bar{x}^{\bar{\mu}}}{\partial x^\mu}  \frac{\partial \bar{x}^{\bar{\nu}}}{\partial x^\nu}   \bar{\Gamma}{}^{\bar{\alpha}} {}_{\bar{\mu}\bar{\nu}}  +
  \frac{\partial x^\alf}{\partial \bar{x}{}^{\bar{\lambda}}} \frac{\partial}{\partial x^\mu} \l(\frac{\partial \bar{x}{}^{\bar{\lambda}}}{\partial x^\nu} \r)\,.
\eea
We define such a set as a "gauge". Usually, the case of zero STC is called as ``coincident gauge'' \cite{Adak:2011ltj,BeltranJimenez:2022azb}. Here, we note again, that in the therm ``coincident gauge'' we mean the only case of coordinates in which the STC is zero (the same as ``Wietzenb\"ock gauge'' in TEGR); and when here we say ``gauge'' we mean the set of all possible coordinates and values of  STCs in them, such that relation (\ref{affconntransf}) is satisfied for each two pairs (${x}^{\mu}$,  $\Gamma {}^\alpha {}_{\mu\nu}$) of the class.

\subsection{Defining the connection: ``turning off'' gravity principle}
\m{turning_off}

Let us repeat the main claims derived above. Teleparallel connections in TEGR and STEGR are not dynamical quantities and left undetermined \cite{Golovnev:2017dox,BeltranJimenez:2019tjy} in theories themselves.

To determine the ISC in TEGR for a {\em given solution} the  generalized ``turning off'' gravity principle was introduced by us in \cite{EPT19,EPT_2020}. This principle is based on the assumption that Noether's current and superpotential  are proportional to contortion components $\stackrel{\bullet}{K} {}^{a} {}_{c\mu }$, or, alternatively, $\stackrel{\bullet}{T} {}^{\alf} {}_{\mu \nu}$ or $\stackrel{\bullet}{S} {}_{a} {}^{\mu \nu}$. In absence of gravity both the current and the superpotential  have to vanish that follows from vanishing  $\stackrel{\bullet}{K} {}^{a} {}_{c\mu }$.
Thus, to determine $\stackrel{\bullet}{A} {}^{a} {}_{c\mu }$ according to this requirement we turn to the formula (\ref{K_A_A}) for a given solution.   For a GR solution under consideration it was suggested:

1)  to choose a convenient tetrad and define $\stackrel{\circ}{A} {}^{a} {}_{c\mu }=-h_b{}^\nu \cN_\mu h^a{}_\nu$, see (\ref{A});

2) to construct related curvature of Levi-Civita spin connection $$\cR{}^i{}_{j\mu\nu}=\di_\mu \cA{}^i{}_{j\nu} - \di_\nu \cA{}^i{}_{j\mu} + \cA{}^i{}_{k\mu}\cA{}^k{}_{j\nu} - \cA{}^i{}_{k\nu}\cA{}^k{}_{j\mu};$$

3) to “switch off” gravity solving  the absent gravity equation $\stackrel{\circ}{R} {}^a {}_{b \gamma \delta}=0$  for parameters of the chosen GR solution;

4) to take $\stackrel{\circ}{A} {}^{a} {}_{c\mu }=\stackrel{\bullet}{A} {}^{a} {}_{c\mu }$ for the found parameter values satisfying $\stackrel{\circ}{R} {}^a {}_{b \gamma \delta}=0$.

\medskip

To determine STC in STEGR we use the adapted for STEGR  ``turning off" gravity principle \cite{EPT:2022uij}. This principle is based on the assumption analogical to that in TEGR, that is Noether's current (\ref{totalcurr}) and superpotential (\ref{totalsup})  have to vanish in absence of gravity. This goal is achieved when $Q_{\alpha \mu \nu}$ (the same $L^{\alpha} {}_{\mu \nu}$) and  $\cR{}^{\alpha} {}_{\beta \mu \nu}$  vanish in the absence of gravity too. To find the STC in STEGR for a GR solution under consideration there are the following steps:

1)  to construct related Riemann curvature tensor of the Levi-Civita connection:
 $$
   \cR{}^\alpha{}_{\beta\mu\nu}=\di_\mu \cG{}^\alpha{}_{\beta\nu} - \di_\nu \cG{}^\alpha{}_{\beta\mu} + \cG{}^\alpha{}_{\kappa\mu}\cG{}^\kappa{}_{\beta\nu} - \cG{}^\alpha{}_{\kappa\nu}\cG{}^\kappa{}_{\beta\mu};
 $$

2) to ``switch off” gravity solving the absent gravity equation $\stackrel{\circ}{R} {}^\alpha {}_{\beta \mu \nu}=0$ for parameters of the chosen GR solution;

3) to take $\Gamma {}^{\alpha} {}_{ \mu \nu}=\stackrel{\circ}{\Gamma} {}^{\alpha} {}_{\mu \nu}$ for the found parameter values which satisfy $\stackrel{\circ}{R} {}^\alpha {}_{\beta \mu \nu}=0$.\\
Torsion of the found connection should be zero automatically because we take it from the Levi-Civita connection for some parameter values, and Levi-Civita connection is always symmetric. Curvature of the found connection should be zero too because we found it from the equation $\stackrel{\circ}{R} {}^\alpha {}_{\beta \gamma \delta}=0$.

%\footnote{Howewer, we always need to keep  $ {R} {}^\alpha {}_{\beta \gamma \delta}=0$ and $T {}^\alpha {}_{\beta \gamma}=0$ in the whole procedure, because if the  coordinate transformation contains special parameters by changing which we ``turned off'' gravity,
%and transformed the connection  as
%\begin{equation}
 %   \Gamma {}^{\iota} {}_{\kappa \lambda}=\Gamma' {}^{\mu} {}_{\nu \pi} \frac{\partial x{}^{\iota}}{\partial x'{}^{\mu}}\frac{\partial x'{}^{\nu}}{\partial x{}^{\kappa}}\frac{\partial x'{}^{\pi}}{\partial x{}^{\lambda}} + \frac{\partial^2 x'{}^{\mu}}{\partial x{}^{\kappa} \partial x{}^{\lambda}}\frac{\partial x{}^{\iota}}{\partial x'{}^{\mu}}
%\end{equation}
% the condition  ${R} {}^\alpha {}_{\beta \gamma \delta}=0$ may be violated, because partial derivatives in (\ref{Riemanntensor}) can be calculated as $ \frac{\partial  }{\partial x{}^{\alpha}}=\frac{\partial x'{}^{\lambda}}{\partial x{}^{\alpha}} \frac{\partial  }{\partial x'{}^{\lambda}}$, where $\frac{\partial x'{}^{\lambda}}{\partial x{}^{\alpha}}$ directly depend on these parameters.}

As it was remarked above, on the level of field equations both in TEGR and STEGR teleparallel connections are arbitrary, they only have to be flat and nothing more. Our generalized principle of ``switching off” gravity formulated above for fixation of teleparallel connection for a given solution does not determine gauges in unique way both in TEGR \cite{EKPT_2021,EKPT_2021a} and in STEGR \cite{EPT:2022uij}.  In general,  ``turning off" gravity in TEGR, the result depends on the tetrad that we choose initially. In general, for each specific tetrad where we  ``turn off" gravity we obtain different gauge (pairs of tetrad and ISC). As a rule, these pairs are not connected by (\ref{lroth}) and (\ref{spin_trans}) applied simultaneously. Thus, torsion, contortion and superpotential being expressed in the same tetrad and unified coordinates are completely different in each case. This gives us different values of conserved quantities.\footnote{ But, sometimes, ``turning off" gravity we can obtain the same gauge \cite{EP:2021snt,EP:2022ohe}.}

 The same way, ``turning off" gravity in STEGR, the result depends on the coordinates that we choose from the start to construct STC for a GR solution under consideration.
%For each coordinates where we  ``turn off" gravity we obtain different connection (being transformed to the same coordinates, the components of each connection are completely  different).
Thus, non-metricity (the same,  disformation) components obtained initially in different coordinates, being transformed to unified coordinates, are  different and this gives us different values of conserved quantities in  each case.
%It is also an interesting question about the conditions when we can find the same connection ``turning off" gravity in different coordinates.

Such problems have to be resolved separately for each solution under consideration, where for each task one has to determine a related gauge. This situation is analogous to a bi-metric representation of GR in \cite{KBL_1997} where, for example, to obtain an acceptable value for the mass of black hole one has to introduce a background metric in special appropriate way. Thus, one of the main purposes of this work is to find the gauges in TEGR and STEGR in which we would have physically meaningful results for the concrete solution - plane gravitational wave in vacuum.

\section{Simplest gauges}
\setcounter{equation}{0}

\subsection{Plane gravitational wave}
\m{App}

Because the wave solutions in GR are  time dependent ones there are no timelike Killing vectors for them. Therefore, in the framework of the fully covariant formalism, first, one does not expect to obtain energy densities related to a fixed observer presenting a fixed frame. Second, because the wave under consideration fills infinite space one cannot obtain any conserved charges (total energy), indeed, in this case one cannot define a fixed observer at infinity or external observer. Therefore, in this paper we do not study a general problem of constructing the energy characteristics of the gravitational wave measured by observers of the aforementioned type. However, of course, one can define proper vectors for freely falling observers, and, correspondingly, to calculate energy-momentum density (current components) measured by such observers. Because such observers should measure zero energy-momentum in correspondence with the equivalence principle the current components have to vanish. By our fully covariant formalism, one has to find gauges which just give zero components of the current in this case following the main task of the paper.

 In this paper, we consider the only plane gravitational wave with only one polarization. The simplest form of the related metric is \cite{Formiga:2018maj}:
\begin{equation}\label{metwave}
g_{\mu\nu} =
\left(
\begin{array}{cccc}
 -1 & 0 & 0 & 0 \\
 0 & [f(t-z)]^2 & 0 & 0 \\
 0 & 0 & [g(t-z)]^2 & 0 \\
 0 & 0 & 0 & 1 \\
\end{array}
\right).
\end{equation}
Here and below the numeration of the coordinates is $t=x^0$, $x=x^1$, $y= x^2$, and $z=x^3$.

In this section, both in TEGR and in STEGR basing on the metric (\ref{metwave}) we construct the simplest gauges. Next, our goal here is to check these simplest gauges both in TEGR and in STEGR on a possibility to achieve a correspondence with the equivalence principle. This goal was initiated, particularly, by the study in \cite{Formiga:2018maj} based on (\ref{metwave}) as well, where such a problem has been considered in another formalism with a subsequent interpretation. We, obtaining similar results, give a different interpretation corresponding to our full covariant formalism.

\subsection{Diagonal polarisation in TEGR}

The simplest tetrad that can be introduced for (\ref{metwave}) is the diagonal tetrad
\begin{equation}\label{tetwave}
 h^A {}_\mu =   \left(
\begin{array}{cccc}
 1 & 0 & 0 & 0 \\
 0 & f(t-z) & 0 & 0 \\
 0 & 0 & g(t-z) & 0 \\
 0 & 0 & 0 & 1 \\
\end{array}
\right).
\end{equation}
To make the formulae shorter, we won't write the argument $(t-z)$ of functions $f$ and $g$ and their derivatives of each order.
Now we follow the ``turning off'' gravity principle in TEGR. For this tetrad, we calculate L-CSC (\ref{A}) which has the non-zero components:
\begin{equation}\label{LCSCsimple}
    \begin{array}{cccc}
    \cA{}^{\hat{0}} {}_{\hat{1} 1} = \cA{}^{\hat{1}} {}_{\hat{0} 1} =-\cA{}^{\hat{1}} {}_{\hat{3} 1}=\cA{}^{\hat{3}} {}_{\hat{1} 1} = f',\\

\cA{}^{\hat{0}} {}_{\hat{2} 2} = \cA{}^{\hat{2}} {}_{\hat{0} 2} =\cA{}^{\hat{3}} {}_{\hat{2} 2} =-\cA{}^{\hat{2}} {}_{\hat{3} 2}=g',\\
    \end{array}
\end{equation}
where prime means a derivative with respect to $u = (t-z)$, and the index with ``hat'' is the tetrad index.
Non-zero components of Riemann tensor constructed with the use of (\ref{LCSCsimple}) are proportional to $ f'' $ or $ g'' $, so turning off gravity means that:
\begin{equation}\label{fg=0}
\begin{array}{cccc}
    f'' =0,  \\
    g'' =0,
\end{array}
\end{equation}
i.e.,
\begin{equation}\label{turningwave}
\begin{array}{cccc}
    f  = {c_1} (t-z)+{d_1},\\
    g = {c_2} (t-z)+{d_2},
\end{array}
\end{equation}
where $c_1$, $c_2$, $d_1$ and $d_2$ are integration constants.
Turning off gravity by (\ref{turningwave}) we get ISC with non-zero components:
\begin{equation}\label{ISCwave}
    \begin{array}{cccc}
    \sA{}^{\hat{0}} {}_{\hat{1} 1} = \sA{}^{\hat{1}} {}_{\hat{0} 1} =-\sA{}^{\hat{1}} {}_{\hat{3} 1}=\sA{}^{\hat{3}} {}_{\hat{1} 1} = {c_1},\\

\sA{}^{\hat{0}} {}_{\hat{2} 2} = \sA{}^{\hat{2}} {}_{\hat{0} 2} =\sA{}^{\hat{3}} {}_{\hat{2} 2} =-\sA{}^{\hat{2}} {}_{\hat{3} 2}= {c_2}.\\
    \end{array}
\end{equation}
Then we calculate superpotential (\ref{super_K}) with (\ref{K_A_A}) definitions which has the non-zero components:
\begin{equation}\label{supdiagpol}
    \begin{array}{cccc}
    \sS{}_{\hat{0}} {}^{0 3} =-\sS{}_{\hat{0}} {}^{3 0}=-\sS{}_{\hat{3}} {}^{0 3}=\sS{}_{\hat{3}} {}^{3 0}= \frac{{c_1}-f'}{f}+\frac{{c_2}-g'}{g},\\
\sS{}_{\hat{1}} {}^{0 1} =\sS{}_{\hat{1}} {}^{3 1} =-\sS{}_{\hat{1}} {}^{1 0} =-\sS{}_{\hat{1}} {}^{1 3} = \frac{{g'}-c_2}{f g},\\
\sS{}_{\hat{2}} {}^{0 2} =\sS{}_{\hat{2}} {}^{3 2} =-\sS{}_{\hat{2}} {}^{2 0} =-\sS{}_{\hat{2}} {}^{2 3} =  \frac{{f'-c_1}}{f g}.\\
    \end{array}
\end{equation}
We assume that superpotential, torsion, etc., should be zero in absence of waves, thus, $c_1=c_2=0$. Then, the ISC (\ref{ISCwave}) becomes
\begin{equation}\label{zeroiscgauge}
    \sA{}^a {}_{b \mu} =0.
\end{equation}
One can classify the pair of the tetrad (\ref{tetwave}) and ISC (\ref{zeroiscgauge}) as a pair of the equivalence class presenting the simplest gauge for the solution (\ref{metwave}) in TEGR.

Now, let us construct the Noether superpotential for this gauge.
First, we consider the simplest case of a freely falling observer which is static in co-moving coordinates of the metric (\ref{metwave}). Then components of his proper vector are:
\begin{equation}\label{proper_FLRW}
   \xi^\sigma=(-1,0,0,0)\,.
\end{equation}
Then  one gets for the  Noether superpotential (\ref{super}) non-zero components:
\begin{equation}
    {\scJ}{}^{0 3} =-{\scJ}{}^{3 0} = \frac{g f'+f g'}{8 \pi }.
\end{equation}
By (\ref{curreqdivsupTEGR}), taking the divergence of superpotential and using the Einstein equations:
\begin{equation}\label{einfg}
    \frac{f''}{f}+\frac{g''}{g}=0
\end{equation}
we get  Noether current
\begin{equation}\label{noethercurrdiag1000}
      {\scJ}{}^{\mu} =  \left\{-\frac{f' g'}{4 \pi },0,0,-\frac{f' g'}{4 \pi }\right\}.
\end{equation}

Let us generalize (\ref{proper_FLRW}). Directly solving the geodesic equation
we derive general freely falling observer's 4-velocity $\xi^\mu$:
\begin{equation}\label{genobs}
    \begin{array}{cccc}
       \xi^0 = - \frac{{C_1}^2 e^{\alpha_0}}{2 f^2}- \frac{{C_2}^2 e^{\alpha_0}}{2 g^2}-\cosh {\alpha_0},
       \\
       \xi^1 = -\frac{{C_1}}{f^2},
       \\
       \xi^2 =- \frac{{C_2}}{g^2},
       \\
       \xi^3 =   -  \frac{{C_1}^2 e^{\alpha_0}}{2 f^2}- \frac{{C_2}^2 e^{\alpha_0}}{2 g^2}-\sinh {\alpha_0},
    \end{array}
\end{equation}
where $C_1$, $C_2$,  $\alpha_0$ are constants of integration. One can see that components (\ref{genobs}) go to the ones in (\ref{proper_FLRW}) when $C_1=C_2=\alpha_0 =0$.

Taking this general form of observer's proper  vector one gets Noether superpotential non-zero components:
\begin{equation}
    \begin{array}{cccc}
  {\scJ}{}^{0 1} ={\scJ}{}^{3 1} =-{\scJ}{}^{1 0} =-{\scJ}{}^{1 3} =  -\frac{{C_1} g'}{8 \pi  f},
\\
{\scJ}{}^{0 2} = {\scJ}{}^{3 2} =-{\scJ}{}^{2 0} =-{\scJ}{}^{2 3} =  -\frac{{C_2} f'}{8 \pi  g},
\\
{\scJ}{}^{0 3} = -{\scJ}{}^{3 0} = \frac{e^{-\alpha_0} \left(g f'+f g'\right)}{8 \pi }.
    \end{array}
\end{equation}
Taking the divergence of superpotential and using the Einstein equations (\ref{einfg}) we get  Noether current
\begin{equation}\label{noethercurrdiagarb}
{\scJ}{}^{\mu} =
   \left\{-\frac{e^{-\alpha_0} f' g'}{4 \pi },0,0,-\frac{e^{-\alpha_0} f' g'}{4 \pi }\right\}.
\end{equation}
Note, first, that the components of the current (\ref{noethercurrdiagarb}) do not depend on $C_1$ and $C_2$; and second, that for $\alpha_0 = 0$ the components (\ref{noethercurrdiagarb}) coincide with (\ref{noethercurrdiag1000}).

Because we have obtained non-zero components of the current (\ref{noethercurrdiag1000}) and  (\ref{noethercurrdiagarb}) measured by a freely falling observer we did not achieve the settled goal to obtain a correspondence with the equivalence principle. Just the analogous conclusion has been given by other authors, for example, in \cite{Formiga:2018maj}, who studied (\ref{metwave}) in the framework other approaches. But, unlike previous approaches, we interpret the failure by unappropriate gauge that stimulates a search for appropriate gauges what we do in next sections.

\subsection{Diagonal polarisation in STEGR}

Let us turn to STEGR where we follow the ``turning off'' gravity principle as well. The non-zero components of Levi-Civita connection for the metric (\ref{metwave}) are:
\begin{equation}\label{cASTEGR}
    \begin{array}{cccc}
\cG{}^{0} {}_{1 1} = \cG{}^{3} {}_{1 1} = f f',\\

\cG{}^{0} {}_{2 2} = \cG{}^{3} {}_{2 2} = g g',\\

\cG{}^{1} {}_{0 1} = \cG{}^{1} {}_{1 0} = -\cG{}^{1} {}_{1 3} = -\cG{}^{1} {}_{3 1} = \frac{f'}{f},\\
\cG{}^{2} {}_{0 2} = \cG{}^{2} {}_{2 0} =-\cG{}^{2} {}_{2 3}=-\cG{}^{2} {}_{3 2}= \frac{g'}{g}.\\
  \end{array}
\end{equation}
In order to switch off gravity one has to equalize the Riemannian tensor with (\ref{cASTEGR}) to zero. By (3.5) the Riemann tensor becomes zero and the Levi-Civita connection has to be taken as the STC (flat and torsionless) which has the non-zero components:
\begin{equation}\label{Connwave}
    \begin{array}{cccc}
 \Gamma{}^{0} {}_{1 1} = \Gamma{}^{3} {}_{1 1} = {c_1} ({c_1} (t-z)+{d_1}),\\

\Gamma{}^{0} {}_{2 2} = \Gamma{}^{3} {}_{2 2} = {c_2} ({c_2} (t-z)+{d_2}),\\

\Gamma{}^{1} {}_{0 1} =\Gamma{}^{1} {}_{1 0} =-\Gamma{}^{1} {}_{1 3} =-\Gamma{}^{1} {}_{3 1} = \frac{{c_1}}{{c_1} (t-z)+{d_1}},\\
\Gamma{}^{2} {}_{0 2}  =-\Gamma{}^{2} {}_{2 3} =-\Gamma{}^{2} {}_{3 2}  =\Gamma{}^{2} {}_{2 0} = \frac{{c_2}}{{c_2} (t-z)+{d_2}}.\\
  \end{array}
\end{equation}
Then we can obtain the non-metricity (\ref{defQ}) and disformation (\ref{defL}).

Non-metricity (\ref{defQ})  in absence of waves (i.e., for Minkowski metric) is:
\begin{equation}
    \begin{array}{cccc}
   Q{}_{0 1 1} =-Q{}_{3 1 1}= -\frac{2 {c_1}}{{c_1} (t-z)+{d_1}},\\

Q{}_{0 2 2} =-Q{}_{3 2 2}= -\frac{2 {c_2}}{{c_2} (t-z)+{d_2}},\\

Q{}_{1 0 1} = Q{}_{1 1 0} = -Q{}_{1 1 3}= -Q{}_{1 3 1}=-\frac{{c_1}}{{c_1} (t-z)+{d_1}}+{c_1} ({c_1} (t-z)+{d_1}),\\

Q{}_{2 0 2} =

Q{}_{2 2 0} = -Q{}_{2 2 3} =-Q{}_{2 3 2} = -\frac{{c_2}}{{c_2} (t-z)+{d_2}}+{c_2} ({c_2} (t-z)+{d_2}).\\
 \end{array}
\end{equation}
We assume that non-metricity should be zero  in absence of waves.
Thus, we have $c_1=c_2=0$ and (\ref{Connwave}) becomes
\begin{equation}\label{zeroconngauge}
    \Gamma{}^{\alpha} {}_{\mu\nu}=0.
\end{equation}
Thus, one can classify the pair of coordinates $(t,x,y,z)$ in the metric (\ref{metwave}) and the STC (\ref{zeroconngauge}) as a pair of the equivalence class  which is the simplest gauge for the solution (\ref{metwave}) in STEGR.

Proceeding further in this gauge  to get the Noether superpotential we choose the observer's proper vector as (\ref{proper_FLRW}).
Komar superpotential (\ref{Komarsup}) becomes zero.
After calculation of the divergent part of the superpotential (\ref{addsupstegr}) we get the total superpotential with non-zero components:
\begin{equation}
   {\cal J}{}^{0 3} =-{\cal J}{}^{3 0} =   {\cal J}_{div}^{0 3} = -{\cal J }_{div}^{3 0} =\frac{  { f'}{g}+{ g'}{f}}{8 \pi }.
\end{equation}
Taking the divergence of the Noether superpotential and using the Einstein equations (\ref{einfg}) we get the Noether current
\begin{equation}\label{noethercurrdiag1000ST}
{\cal J}{}^{\mu} =
    \left\{-\frac{f' g'}{4 \pi },0,0,-\frac{ f' g'}{4 \pi}\right\}.
\end{equation}
Note that it exactly coincides with (\ref{noethercurrdiag1000}).

Second, to get another Noether superpotential we choose the observer's proper vector as in (\ref{genobs}).
Komar superpotential (\ref{Komarsup}) is zero again.
Calculating the divergent part  (\ref{addsupstegr}) we get the total superpotential
\begin{equation}
    \begin{array}{cccc}
{\cal J}{}^{0 1} ={\cal J}{}^{3 1}  =-{\cal J}{}^{1 0}  =-{\cal J}{}^{1 3} =  -{C_1}  \frac{{ f'}{g}+{ g'}{f}}{8 \pi  f^2},
\\
{\cal J}{}^{0 2} ={\cal J}{}^{3 2} = -{\cal J}{}^{2 0}= -{\cal J}{}^{2 3}= -{C_2} \frac{{ f'}{g}+{ g'}{f}}{8 \pi  g^2},
\\
{\cal J}{}^{0 3} = -{\cal J}{}^{3 0} =\frac{e^{-\alpha_0}  \left({ f'}{g}+{ g'}{f}\right)}{8 \pi }.
     \end{array}
\end{equation}
And then taking the divergence  and using the Einstein equations (\ref{einfg}) we get the Noether current
\begin{equation}\label{noethercurrdiagarbST}
{\cal J}{}^{\mu} =
   \left\{-\frac{e^{-\alpha_0} f' g'}{4 \pi },0,0,-\frac{e^{-\alpha_0} f' g'}{4 \pi }\right\}.
\end{equation}
Note that it exactly coincides with (\ref{noethercurrdiagarb}).

Again, obtaining non-zero components of the current (\ref{noethercurrdiag1000ST}) and  (\ref{noethercurrdiagarbST}) measured by a freely falling observer, one concludes that the settled goal to obtain a correspondence with the equivalence principle is not achieved in this case. Again, we interpret the failure by unappropriate gauge of coordinates $(t,x,y,z)$ in the metric (\ref{metwave}) and the STC (\ref{zeroconngauge})  that stimulates a search for appropriate gauges that we do in next sections.

\subsection{A linear approximation}

In this subsection, we consider a linear approximation of the results in previous two subsections. It is initiated by two reasons. First, a linear approximation has already  been suggested in other works in other formalisms, see \cite{Maluf:2003fg,Formiga:2018maj,Formiga_2020}, where a correspondence with the equivalence principle is absent as well. Therefore, it is useful to give a comparison of these results with ours. Second, in the works \cite{Maluf:2003fg,Formiga:2018maj,Formiga_2020} there is a coincidence of the linear approximation with the Landau-Lifshitz energy-momentum density for the linear gravitational wave \cite{Landau_Lifshitz_1975}. Therefore, it is quite desirable to provide a linear approximation of our results and give an appropriate interpretation.

Now let us represent the Noether currents both in
TEGR (\ref{noethercurrdiag1000}) and in STEGR (\ref{noethercurrdiag1000ST}), which are the same, in the linear approximation. For the weak wave in Minkowski space we
choose the simplest presentation \cite{Landau_Lifshitz_1975}:
\begin{equation}\label{metlin}
   g_{\mu\nu}= \eta_{\mu\nu}+ h_{\mu\nu},
\end{equation}
where $\eta_{\mu\nu} = \diag [-1,~+1,~+1,~+1]$. Thus, in the case of the metric (\ref{metwave}) in TT-gauge \cite{Landau_Lifshitz_1975} one has
\begin{equation}\label{metlinfg}
    f=1+\frac{1}{2} h_{xx} (t-z),~~g=1-\frac{1}{2} h_{xx} (t-z).
\end{equation}
The main item in calculation of the current is a determined gauge. Turning to the exact consideration we have defined such gauges, they are the tetrad (\ref{tetwave}) and ISC (\ref{zeroiscgauge}) in TEGR, and coordinates $(t,x,y,z)$ in the metric (\ref{metwave}) and the STC (\ref{zeroconngauge}) in STEGR. By the logic of construction, to represent these gauges in the linear approximation one has to only exchange the exact metric (\ref{metwave}) in expressions (including a tetrad (\ref{tetwave})) by the linearized metric (\ref{metlinfg}).
%Switching off gravity in the linear case gives the same zero ISC (\ref{zeroiscgauge}) and zero STC (\ref{zeroconngauge}).
Switching off gravity in TEGR and STEGR in the linear case directly we first equate the Riemann tensor of the metric (3.1) with the condition (3.25) to zero and find that  $h''_{xx}=0$. Integrating it we get the condition $h_{xx} = c_1 (t-z) + d_1$. Calculating superpotential and the non-metricity with the latter we find again that these tensors are zero when $c_1=0$ and, thus, the corresponding ISC and STC are zero. So, the linearization procedure and the "turning off" gravity procedure commute here.

Another necessary item  in calculation of the current is a determined displaced vector $\xi^\sig$. For a spacetime with the metric (\ref{metlin}) one obtains for a freely falling observer:
\begin{equation}\label{xi_lin}
   \xi^\sig = \l(-1,0,0,0 \r)\,
\end{equation}
that coincides with (\ref{proper_FLRW}) for the exact metric. Taking into account all the above, the current (\ref{noethercurrdiag1000}), the same (\ref{noethercurrdiag1000ST}), becomes in the lower order
\begin{equation}\label{noethercurrdiag1000lin}
{\cal J}{}^{\mu}_{lin} =
   \left\{ \frac{{h'}_{xx}^2}{16 \pi },~0,~0,~ \frac{{h'}_{xx}^2}{16 \pi }\right\}.
\end{equation}
Considering a more general case of freely falling observers (\ref{genobs}) in the
linear approximation (\ref{metlin}) and (\ref{metlinfg}), one obtains for the current (\ref{noethercurrdiagarb}), the same (\ref{noethercurrdiagarbST}) analogous expression
\begin{equation}\label{noethercurrdiagarblin}
    {\cal J}{}^{\mu}_{lin} =
   \left\{\frac{e^{-\alpha_0}{h'}_{xx}^2}{16 \pi },~0,~0,~\frac{e^{-\alpha_0}{h'}_{xx}^2}{16 \pi }\right\},
\end{equation}
Because we see nonzero components for the currents we have no a correspondence with the equivalence principle in linear approximation as well, that is not surprisingly due to the exact consideration.
The discrepancy in interpretation of the equivalence principle in the framework of our approach in linear approximation is explained in the same way as in the exact case, namely, the used gauges are not appropriate ones.

However, let us return to the expression (\ref{noethercurrdiag1000lin}) and concentrate to it.  It exactly coincides with the result in \cite{Landau_Lifshitz_1975} for the case of the weak plane gravitational wave with only one polarization propagating on a flat background.
The component  ${\cal J}{}^{0}_{lin}$ represents the energy density, whereas  ${\cal J}{}^{3}_{lin}$ - the energy density flux of such a wave propagating along $z$ axis. One can see that  these quantities are positively defined. The same coincidence has been obtained for the gravitational wave
 (\ref{metwave}) in other formalisms, for example,  in \cite{Formiga:2018maj}. This fact is considered there as a criterion that supports correctness of the result. In the following paper \cite{Formiga_2020}, this coincidence is supported by a relation to a so-called “ideal frame”.

To be more convincible we need to explain the quite acceptable result (\ref{noethercurrdiag1000lin}), where components are explained as energetic characteristics of
gravitational wave, in the framework of our formalism. Let us consider deeper the Landau-Lifshitz prescription, which is based on their pseudotensor. Like all the pseudotensors, it is non-covariant --– it is not transformed as
tensor under arbitrary coordinate transformations what makes the physical interpretation of the
conserved quantities more difficult. One of the ways to improve the situation is a possibility to
covariantize pseudotensors by introducing a fixed Minkowskian background with the Minkowski metric (together with the dynamical spacetime with the metric
tensor $g_{\mu\nu}$), see book
\cite{Petrov_KLT_2017}. Then, the Landau-Lifshitz conserved pseudotensor $t^{\mu\nu}_{LL}$, which has the
mathematical weight +2, permits to construct us a conserved covariant current\footnote{The analogous method of covariantization can be applied to an arbitrary pseudotensor.}
\begin{equation}\label{Jll}
     {\cal J}{}^{\mu}_{LL} = \left(\sqrt{-\det \eta_{\alpha\beta}} \right)^{-1} t^{\mu\nu}_{LL} \bar{\xi} {}_\nu .
\end{equation}
From the beginning it is assumed that (\ref{Jll}) is written in the Lorentzian coordinates, of course, $-\det \eta_{\alpha\beta}=1$. Then, ${\cal J}{}^{\mu}_{LL}$ is thought as a vector density of the weight +1 and can be represented in arbitrary coordinates by the ordinary way if partial derivatives in $t^{\mu\nu}_{LL}$ are replaced by covariant ones. Vector $\bar{\xi} {}^\mu$ is a Killing vector of the Minkowski space.

Let us choose
\begin{equation}\label{killlin}
    \bar{\xi} {}^\mu = (-1,~0,~0,~0),
\end{equation}
as a timelike Killing vector of the Minkowski space. Then the current (\ref{Jll}) becomes
\begin{equation}\label{Jll+}
     {\cal J}{}^{\mu}_{LL} = (~ t^{00}_{LL} ,~ t^{10}_{LL} ,~ t^{20}_{LL} ,~ t^{30}_{LL} ) \,.
\end{equation}
Thus, the interpretation of  ${\cal J}{}^{0}_{LL}$ and ${\cal J}{}^{3}_{LL}$  as energy density and energy density flux coincides with a related
interpretation in $t^{00}_{LL}$ and $t^{30}_{LL}$ in the book \cite{Landau_Lifshitz_1975}. In linear approximation, the expression (\ref{Jll+}) coincides with (\ref{noethercurrdiag1000lin}).

Return again to the decompositions (\ref{metlin}) and (\ref{metlinfg}), where perturbations are considered on a flat background with the Minkowski metric. Now, let us exchange interpretation. Assume that in the total current  (\ref{noethercurrdiag1000}), the same (\ref{noethercurrdiag1000ST}), one uses the timelike Killing vector of Minkowski space (\ref{killlin}) instead of the freely falling observer proper vector (\ref{proper_FLRW}) (although, formally they are the same). As a result,  one, of course, obtains again (\ref{noethercurrdiag1000lin}) after linearization. However, the coincidence of the result (\ref{noethercurrdiag1000lin}) with the Landau-Lifshitz one becomes clear.

Let us consider (\ref{Jll}) without an approximation. Thus,
$t^{\mu\nu}_{LL}$ for the metric (\ref{metwave}) is
\begin{equation}\label{Jllstrong}
  {\cal J}{}^{\mu}_{LL} =  \left\{ -\frac{4 f g f' g'+g^2 f'^2+f^2 g'^2}{8 \pi },~0,~0,~ -\frac{4 f g f' g'+g^2 f'^2+f^2 g'^2}{8 \pi }\right\}.
\end{equation}
One can see that it differs from (\ref{noethercurrdiag1000}) and (\ref{noethercurrdiag1000ST}), although its linear approximation gives (\ref{noethercurrdiag1000lin}). We stress that our results (\ref{noethercurrdiag1000}) and (\ref{noethercurrdiag1000ST}) are obtained in the framework of the initially covariant method, whereas (\ref{Jllstrong}) is obtained after covariantization (an additional procedure) of the Landau-Lifshitz pseudotensor.

               Finally, it is interesting to discuss the results (\ref{noethercurrdiagarb}), (\ref{noethercurrdiagarbST}) and (\ref{noethercurrdiagarblin}). The vector (\ref{genobs})  is not a Killing vector of Minkowski space. However, without changing the results (\ref{noethercurrdiagarb}), (\ref{noethercurrdiagarbST}) and (\ref{noethercurrdiagarblin}) one can set $C_1 = C_2 = 0$. By this, we can claim that we use vector
\begin{equation}\label{killingalpha}
    \xi^\mu_{boosted} = \left\{-\cosh \alpha_0,~0,~0,~-\sinh \alpha_0\right\},
\end{equation}
which is a timelike Killing vector of Minkowski space. Indeed, boosting (\ref{killlin}) by global Lorentz transformation in Minkowski space one gets (\ref{killingalpha}). This transformation has a form: $\xi^\mu_{boosted} =\lambda {}^\mu {}_\nu \bar\xi^\nu$, where
\begin{equation}\label{lambdaboost}
  \lambda {}^\mu {}_\nu=
\left(
\begin{array}{cccc}
  \cosh \alpha_0   & 0 & 0 & \sinh \alpha_0 \\
  0 & 0 & 0 & 0\\
    0 & 0 & 0 & 0\\
  \sinh \alpha_0    & 0 & 0 & \cosh \alpha_0
\end{array}
\right) =
\left(
\begin{array}{cccc}
  \frac{1}{\sqrt{1-v^2}}  & 0 & 0 &   \frac{v}{\sqrt{1-v^2}} \\
  0 & 0 & 0 & 0\\
    0 & 0 & 0 & 0\\
    \frac{v}{\sqrt{1-v^2}}   & 0 & 0 &   \frac{1}{\sqrt{1-v^2}}
\end{array}
\right)
\end{equation}
with $v$ a 3-dimensional constant velocity in Minkowski spacetime.
Then interpretation of (\ref{noethercurrdiagarb}), (\ref{noethercurrdiagarbST}) and (\ref{noethercurrdiagarblin}) becomes clear: they are obtained by boosting the vector $\xi^\mu$ by (\ref{lambdaboost}) in  (\ref{noethercurrdiag1000}), (\ref{noethercurrdiag1000ST}) and (\ref{noethercurrdiag1000lin}).

The result of linear approximation is an important result not only because it coincides with the Landau-Lifshitz prediction (and other pseudotensor approaches as well), but mainly because it is checked observationally. This possibly means that the equivalence principle itself may not require necessary zero current
 \textit{in this concrete case}, and non-zero values can have a physical meaning with an appropriate interpretation. On the other hand, due to lack of observational evidence for a strong gravitational wave regime, we can not a priory say if the general form of the current has the same meaning, regarding its difference from the pseudo-tensor result for a strong wave.

\section{Gauges compatible with the equivalence principle}
\setcounter{equation}{0}

In this section, as it was announced above, we are searching for gauges which give zero  current for a free falling observer. Zero values of the Noether current components measured by such observers mean that they detect the absence of gravity, that is a correspondence  with the equivalence principle. In another word, we are searching for gauges compatible with the equivalence principle.

\subsection{Gauge changing in TEGR}

Let us try to obtain zero Noether current in TEGR by changing a gauge. To do this, we change the ISC as usual (\ref{spin_trans}):
\begin{equation}\label{lrotA}
{\sA}{}'^a {}_{b \mu} = \Lambda^a {}_c (x^\nu) {\sA}{}^c {}_{d \mu} \Lambda^b {}_d (x^\nu) + \Lambda^a {}_c(x^\nu) \partial_{\mu} \Lambda_b {}^c (x^\nu)\,.
\end{equation}
Because in the previous section we had zero ISC (\ref{zeroiscgauge}), our new ISC has a form (\ref{telcon})
\begin{equation}\label{lrotA0}
{\sA}{}'^a {}_{b \mu} = \Lambda^a {}_c(x^\nu) \partial_{\mu} \Lambda_b {}^c (x^\nu)\,,
\end{equation}
while the tetrad remains the same (\ref{tetwave}).
Or, another way is as follows: the tetrad changes as (\ref{lroth})
\begin{equation}\label{lroth0}
h'^a {}_{\mu} = \Lambda^a {}_b (x^\nu) h^b {}_\mu\,,
\end{equation}
while  ISC (\ref{zeroiscgauge}) remains zero.
In this section, we also assume that the affine connection $\sG {}^\alf {}_{\mu\nu}$ should have the same symmetries as a solution (\ref{metwave}). This means that displacements in $x$, in $y$ and in $t+z$ directions do not change it. Such a proposal was applied to the affine connection in modified teleparallel theories with other symmetries in \cite{Hohmann:2019nat}. Because in modified teleparallel theories the connection is dynamical, both the metric and the connection should have the same symmetries as solution. In TEGR, the ISC and affine connection are non dynamical, so, at the level of field equations, the requirement for them to have the same symmetries as metric might be too strong. Nevertheless, we consider Noether's current (\ref{noethercurrtegr}) and superpotential (\ref{super}) and can require for them the same symmetries as metric has. For this purpose it is sufficient to require the same symmetries for the teleparallel superpotential, contortion or  torsion. For example we can require for contortion:
\begin{equation}
    {\pounds}_\xi \sK {}^\alf {}_{\mu\nu}=0,
\end{equation}
where ${\pounds}_\xi$ is Lie derivative and $\xi$ is a Killing vector of the solution. For the metric (\ref{metwave}), the symmetry holds along the directions $\Delta x$: $\xi^\mu= (0,~1,~0,~0)$,
$\Delta y$: $\xi^\mu=(0,~0,~1,~0)$, and $\Delta(t+z)$: $\xi^\mu=(1,~0,~0,~1).$
The contortion is
\begin{equation}
     \sK {}^\alf {}_{\mu\nu}= \sG {}^\alf {}_{\mu\nu}- \cG {}^\alf {}_{\mu\nu}=0,
\end{equation}
where   the Levi-Civita connection has the same symmetries as metric (\ref{metwave}). Now let's go back to definitions (\ref{ISCdef}) and (\ref{telcon}). The tetrad (\ref{tetwave}) is already symmetric as metric (\ref{metwave}). Therefore, $\sG {}^\alf {}_{\mu\nu}$ is symmetric when the ISC (\ref{telcon}) is symmetric.
This requirement is fulfilled when $ \Lambda^a {}_b (x^\nu)=\Lambda^a {}_b (t-z)$ in (\ref{telcon}), that is an arbitrary Lorentz rotation depends on $t-z$ only.

Keeping in mind matrices of local Lorentz rotations obtained by this prescription and given in Appendix A,  we have found that for the related gauges (pairs of tetrad (\ref{tetwave}) and ISC (\ref{telcon}), where  $\Lambda^a {}_b (t-z)$ is a composition of matrices given in Appendix A, or transformed tetrad (\ref{tetwave}) by $\Lambda^a {}_b (t-z) h^b {}_\mu$ and zero ISC (\ref{zeroiscgauge})) the Noether current does not change! Thus one cannot find a gauge for which the current vanishes for a free moving observer with  such   a restrictive condition.

Now let's assume that $\Lambda {}^a {}_b$ can be non-symmetrical, for example, along the direction $dx$, thus, it can depend  on $x$ and $(t-z)$. The motivation for this proposal is that the connection in TEGR  is not dynamical and thus cannot be felt by observers like symmetrical metric can, and, thus, we don't need $\Lambda {}^a {}_b$ to be symmetrical. When we assume the dependence on $x$, the Noether superpotential can depend on $x$.
However, making the Noether current be equal to zero, such a current will automatically satisfy the symmetries of the solution.
One of the most simple Lorentz rotations $\Lambda {}^a {}_b$ which depend on $x$ and $(t-z)$ and can change the Noether current is
\begin{equation}\label{lamalfx}
  \Lambda {}^a {}_b=  \left(
\begin{array}{cccc}
 1 & 0 & 0 & 0 \\
 0 & \cos (x {\psi}(t-z)) & 0 & \sin (x {\psi}(t-z)) \\
 0 & 0 & 1 & 0 \\
 0 & -\sin (x {\psi}(t-z)) & 0 & \cos (x {\psi}(t-z)) \\
\end{array}
\right).
\end{equation}
The ISC (\ref{telcon}) $\sA{}^a {}_{c \mu} = \Lambda {}_b {}^c \partial_\mu \Lambda {}^a {}_b$ calculated with (\ref{lamalfx}) is
\begin{equation}\label{ISCalfx}
\begin{array}{cccc}
\sA{}^{\hat{1}} {}_{\hat{3} 1} =-\sA{}^{\hat{3}} {}_{\hat{1} 1}= {\psi}(t-z);\\
       \sA{}^{\hat{1}} {}_{\hat{3} 0} = \sA{}^{\hat{3}} {}_{\hat{1} 3} = -\sA{}^{\hat{1}} {}_{\hat{3} 3}= -\sA{}^{\hat{3}} {}_{\hat{1} 0}= x {\psi}'(t-z).\\
\end{array}
\end{equation}
Then one can calculate the contortion (\ref{K_A_A}) with (\ref{lamalfx}) and (\ref{LCSCsimple}). Then, the teleparallel
superpotential (\ref{super_K}) is
\begin{equation}\begin{array}{cccc}
  \sS{}_{\hat{0}} {}^{0 1} =
\sS{}_{\hat{0}} {}^{3 1} =
-\sS{}_{\hat{0}} {}^{1 0} =
-\sS{}_{\hat{0}} {}^{1 3} =
-\frac{x {\psi}'(t-z)}{f};
\\
\sS{}_{\hat{0}} {}^{0 3} =
-\sS{}_{\hat{0}} {}^{3 0} =
-\frac{f'+{\psi}(t-z)}{f}-\frac{g'}{g};
\\
\sS{}_{\hat{1}} {}^{0 1} =
\sS{}_{\hat{1}} {}^{3 1} =
-\sS{}_{\hat{1}} {}^{1 0} =
-\sS{}_{\hat{1}} {}^{1 3} =
\frac{g'}{f g};
\\
\sS{}_{\hat{2}} {}^{0 2} =
-\sS{}_{\hat{2}} {}^{2 0} =
\frac{f'}{f g};
\\
\sS{}_{\hat{2}} {}^{1 2} =
-\sS{}_{\hat{2}} {}^{2 1} =
\frac{x {\psi}'(t-z)}{f g};
\\
\sS{}_{\hat{2}} {}^{2 3} =
-\sS{}_{\hat{2}} {}^{3 2} =
-\frac{f'+{\psi}(t-z)}{f g};
\\
\sS{}_{\hat{3}} {}^{0 3} =
-\sS{}_{\hat{3}} {}^{3 0} =
\frac{f'}{f}+\frac{g'}{g}.
\end{array}
\end{equation}
Taking (\ref{proper_FLRW}) we have the Noether superpotential  in TEGR (\ref{super}):
\begin{equation}\label{netsupalftegr}
\begin{array}{cccc}
 {\scJ}{}^{0 1} = {\scJ}{}^{3 1} =-{\scJ}{}^{1 0}=-{\scJ}{}^{13}= \frac{x g {\psi}'(t-z)}{8 \pi };
\\
{\scJ}{}^{0 3} = -{\scJ}{}^{3 0}=\frac{g \left(f'+{\psi}(t-z)\right)+f g'}{8 \pi };
\end{array}
\end{equation}
Then, taking the divergence (\ref{curreqdivsupTEGR}) of (\ref{netsupalftegr}) we get the Noether current in TEGR:
\begin{equation}
 {\scJ}{}^{\mu} = \left\{-\frac{g f''+2 f' g'+f g''+ {\psi}(t-z) g'}{8 \pi },~0,~0,~-\frac{g f''+2 f' g'+f g''+ {\psi}(t-z) g'}{8 \pi }  \right\}.
\end{equation}
Applying here the Einstein equation (\ref{einfg}) we get:
\begin{equation}\label{netcurralphagauge}
 {\scJ}{}^{\mu}= \left\{ -\frac{2 f' g'+ {\psi}(t-z) g'}{8 \pi },~0,~0,~ -\frac{2 f' g'+ {\psi}(t-z) g'}{8 \pi }  \right\}.
\end{equation}
Then, the condition for zero Noether current is
\begin{equation}
   {\psi}(t-z)= -2 f' .
\end{equation}

Thus a gauge compatible  with the equivalence principle is constructed. Analogously, one can permit a  dependence on $y$ and $(t-z)$ with the same result. More complicated gauge constructed with the local Lorentz rotations depending simultaneously on $x$, $y$ and $(t-z)$ is considered in the next section on the basis of work \cite{Obukhov:2009gv}.

\subsection{Gauge changing in STEGR}

In this subsection, from the start we restrict ourselves by the simplest requirement as well.  We assume that changed Noether conserved quantities have to depend on $(t-z)$ only.
Komar superpotential obtained for the metric (\ref{metwave}) and vector (\ref{proper_FLRW}) is left zero independently on transformations which change a gauge. Now, consider the additional part (\ref{addsupstegr}).
Because the metric (\ref{metwave}) depend on $(t-z)$ only,  the STC  $\Gamma {}^\alf {}_{\mu\nu}$ included into the additional part of Noether superpotential (\ref{addsupstegr}) the non-metricity (\ref{defQ}) should depend on $(t-z)$ only too. Then we check if such Noether superpotentials (depending on $(t-z)$ only) can satisfy the equivalence principle.

We assume that there exist some new  coordinates $(T,X,Y,Z)$ in which the STC $\Gamma {}^\alf {}_{\mu\nu}=0$.
New  coordinates $(T,X,Y,Z)$ depend on the  coordinates $(t,x,y,z)$ in a general way as:
\begin{equation}\label{Xx}
 T=t+ \Delta T,
   ~~  X=x+\Delta X,
  ~~  Y=y+\Delta Y,
  ~~  Z=z+\Delta Z.
\end{equation}
When (\ref{Xx}) is expanded in a Taylor series, the functions $\Delta T$,  $\Delta X$,  $\Delta Y$,  $\Delta Z$,  depend on the derivatives of $(T,X,Y,Z)$ with respect  to $(t,x,y,z)$. These derivatives are included in the formula of the transformed STEGR flat connection, which (after applying the coordinate transformation from $(T,X,Y,Z)$ to  $(t,x,y,z)$) is calculated as
\begin{equation}\label{GammaXx}
\Gamma {}^\alf {}_{\mu\nu}=    \frac{\partial x^\alf}{\partial X^\lambda} \frac{\partial}{\partial x^\mu} \l(\frac{\partial X^\lambda}{\partial x^\nu} \r),
\end{equation}
where $x^\mu \equiv (t,x,y,z)$, $X^\mu \equiv (T,X,Y,Z)$.
To make the STC (\ref{GammaXx}) depending only on $(t-z)$,
the derivatives of $(T,X,Y,Z)$ with respect  to $(t,x,y,z)$ and, thus, it is sufficient to make the functions $\Delta T$,  $\Delta X$,  $\Delta Y$,  $\Delta Z$ be  dependent on $(t-z)$  only. Thus,
  assuming that the functions $\Delta T$,  $\Delta X$,  $\Delta Y$,  $\Delta Z$ depend on $(t-z)$ only we get for the STC (\ref{GammaXx}) non-zero components
\begin{equation}
    \begin{array}{cccc}
      \Gamma{}^{0} {}_{ 0 0} =
\Gamma{}^{0} {}_{ 3 3} =
-\Gamma{}^{0} {}_{ 0 3} =
-\Gamma{}^{0} {}_{ 3 0} =
\frac{{\Delta T}' {\Delta Z}''-{\Delta T}'' \left({\Delta Z}'-1\right)}{{\Delta T}'-{\Delta Z}'+1};
\\

\Gamma{}^{1} {}_{ 0 0} =
\Gamma{}^{1} {}_{ 3 3} =
\frac{{\Delta X}' \left({\Delta Z}''-{\Delta T}''\right)}{{\Delta T}'-{\Delta Z}'+1}+{\Delta X}'';
\\
\Gamma{}^{1} {}_{ 0 3} =
\Gamma{}^{1} {}_{ 3 0} =
\frac{{\Delta X}' \left({\Delta T}''-{\Delta Z}''\right)+{\Delta X}'' \left(-{\Delta T}'+{\Delta Z}'-1\right)}{{\Delta T}'-{\Delta Z}'+1};
\\
\Gamma{}^{2} {}_{ 0 0} =
\Gamma{}^{2} {}_{ 3 3} =
\frac{{\Delta Y}' \left({\Delta Z}''-{\Delta T}''\right)}{{\Delta T}'-{\Delta Z}'+1}+{\Delta Y}'';
\\
\Gamma{}^{2} {}_{ 0 3} =
\Gamma{}^{2} {}_{ 3 0} =
\frac{{\Delta Y}' \left({\Delta T}''-{\Delta Z}''\right)+{\Delta Y}'' \left(-{\Delta T}'+{\Delta Z}'-1\right)}{{\Delta T}'-{\Delta Z}'+1};
\\
\Gamma{}^{3} {}_{ 0 0} =
\Gamma{}^{3} {}_{ 3 3} =
-\Gamma{}^{3} {}_{ 0 3} =
-\Gamma{}^{3} {}_{ 3 0} =
\frac{\left({\Delta T}'+1\right) {\Delta Z}''-{\Delta T}'' {\Delta Z}'}{{\Delta T}'-{\Delta Z}'+1},
    \end{array}
\end{equation}
where the functions $\Delta T$,  $\Delta X$,  $\Delta Y$,  $\Delta Z$, depend on $u=(t-z)$ only, and prime means the differentiation with respect to $u$ again.

We take again the observer's proper vector (\ref{proper_FLRW}).
Because Komar superpotential (\ref{Komarsup}) remains zero the total Noether superpotential (\ref{totalsup}) is determined only by additional part  (\ref{addsupstegr}) which has non-zero components:
\begin{equation}\label{noethersupTXTZcoin}
{\cal J}{}^{0 3} = -{\cal J}{}^{3 0}=\frac{g \left(2 f' \left(\Delta T'-\Delta Z'+1\right)+f \left(\Delta Z''-\Delta T''\right)\right)+2 f g' \left(\Delta T'-\Delta Z'+1\right)}{16 \pi
   \left(\Delta T'-\Delta Z'+1\right)}.
\end{equation}
To make the Noether current zero, it is sufficient to make the Noether superpotential (following (\ref{CL_c_s})) be constant ${\cal J}{}^{0 3} = -{\cal J}{}^{3 0} =A_0$. Then we easily find that
\begin{equation}
\frac{(\Delta T - \Delta Z)''}{(\Delta T - \Delta Z)'} =  2 \left(-\frac{8 \pi  {A_0}}{f g}+\frac{f'}{f}+\frac{g'}{g}\right).
\end{equation}
Then
\begin{equation}\label{TZarb}
(\Delta T - \Delta Z )'  =    {A}_1 f^2   g^2  \exp \left(-\int \left(\frac{16 \pi  {A_0}}{f g}\right) \, du  \right)  ,
\end{equation}
where $A_1$ is a constant of integration.
 If we assume $A_0=0$, in this case the Noether superpotential is zero and
\begin{equation}\label{TZzeronetcur}
    \Delta T = \Delta Z + A_1 \int f^2 g^2 du + A_2,
\end{equation}
where $A_1$ and $A_2$ are constants of integration. One can easily see that
$\Delta Z$, $\Delta X$, $\Delta Y$ can be arbitrary functions. Concerning $\Delta T$, it is connected with $\Delta Z$ by   (\ref{TZarb}) in the case of non-zero constant Noether current or by (\ref{TZzeronetcur}) in the case of zero Noether current.

In STEGR, in contrast to TEGR, even with a very strong restriction on the STC and Noether superpotential, requiring dependence on $t-z$ only, we have reached the goal, that is, we have found a gauge with zero current for a freely moving observer.

\section{Obukhov-Pereira-Rubilar gauge}
\setcounter{equation}{0}

In this section, we continue searching for gauges which give zero  current for a freely falling observer, thus, gauges compatible with the equivalence principle. For this we use the representation of the plane wave solution in a special way. Namely, in \cite{Obukhov:2009gv} and \cite{Maluf:2008yy} the authors study the problem of the energy that can be brought by flat fronted gravitational wave. They use different set of coordinates $(T,X,Y,Z)$, for which in $(-,+,+,+)$ signature the metric for the gravitational wave solution  has a form
\begin{equation}\label{MetWaveOb}
 g_{\mu\nu} =    \left(
\begin{array}{cccc}
- H(T-Z,X,Y)-1 & 0 & 0 & H(T-Z,X,Y) \\
 0 & 1 & 0 & 0 \\
 0 & 0 & 1 & 0 \\
 H(T-Z,X,Y) & 0 & 0 & -H(T-Z,X,Y)+1 \\
\end{array}
\right).
\end{equation}
The Einstein equations in vacuum acquire the simple form:
\begin{equation}\label{EinWaveOb}
    \frac{\partial^2 H(T-Z,X,Y)}{\partial X^2}+\frac{\partial^2 H(T-Z,X,Y)}{\partial Y^2}=0.
\end{equation}
We discuss the results of \cite{Obukhov:2009gv} in Appendix B in detail. Here, we consider the tetrad suggested in \cite{Obukhov:2009gv} in the framework of our formalism. The reason is that result of \cite{Obukhov:2009gv} with zero energetic characteristics  could be interpreted as having a relation to the equivalence principle.  Thus,
Obukhov et. al. \cite{Obukhov:2009gv} use  a tetrad for the metric (\ref{MetWaveOb}):
\begin{equation}\label{TetWaveOb}
 h^a {}_\mu =   \left(
\begin{array}{cccc}
 \frac{1}{2} H(T-Z,X,Y)+1 & 0 & 0 & -\frac{1}{2} H(T-Z,X,Y) \\
 0 & 1 & 0 & 0 \\
 0 & 0 & 1 & 0 \\
 -\frac{1}{2} H(T-Z,X,Y) & 0 & 0 & \frac{1}{2} H(T-Z,X,Y)-1 \\
\end{array}
\right).
\end{equation}

Here, we need to use the coordinates of (\ref{metwave}). To transform the metric (\ref{MetWaveOb}) to the metric (\ref{metwave}) under consideration we use the Formiga's \cite{Formiga_2020} Eq(2.32):
\begin{equation}\label{TXYZtotxyz}
\begin{array}{cccc}
        T=t+\frac{1}{2} \l(x^2 f(U) \frac{df}{dU}+y^2 g(U) \frac{dg}{dU} \r), \\
        Z=z+\frac{1}{2} \l(x^2 f(U) \frac{df}{dU}+y^2 g(U) \frac{dg}{dU} \r), \\
        X=f(U) x,~~     Y=g(U) y\\
        U=T-Z=t-z,\\
         H(U,X,Z)=-\frac{1}{f} \frac{d^2 f}{dU^2} (X^2-Y^2).
\end{array}
\end{equation}

\subsection{Zero current in TEGR}

After the coordinate transformation (\ref{TXYZtotxyz}) $(T,X,Y,Z) \goto (t,x,y,z)$ metric (\ref{MetWaveOb}) transforms to diagonal form (\ref{metwave}).
Tetrad (\ref{TetWaveOb}) transforms to the form
\begin{equation}\label{TetWaveObintxyz}
 h^a {}_\mu =   \left(
\begin{array}{cccc}
 \frac{1}{2} \left(x^2 f'^2+y^2 g'^2+2\right) & x f f' & y g g' & \frac{1}{2} \left(-x^2 f'^2-y^2 g'^2\right) \\
 x f' & f & 0 & -x f' \\
 y g' & 0 & g & -y g' \\
 \frac{1}{2} \left(-x^2 f'^2-y^2 g'^2\right) & -x f f' & -y g g' & \frac{1}{2} \left(x^2 f'^2+y^2 g'^2-2\right) \\
\end{array}
\right).
\end{equation}
As is was considered in \cite{Obukhov:2009gv}, the corresponding ISC to the tetrad (\ref{TetWaveOb}) was zero. Because ISC transforms as spacetime vector under spacetime transformations, the corresponding to (\ref{TetWaveObintxyz}) ISC remains zero. Besides, the tetrad (\ref{TetWaveObintxyz}) is connected to the diagonal tetrad (\ref{tetwave}) by $h^a {}_\mu = \Lambda^a {}_b h_{(diag)}^b{}_\mu $, where the Lorentz rotation $\Lambda^a {}_b$ is
\begin{equation}
\Lambda^a {}_b =    \left(
\begin{array}{cccc}
 \frac{1}{2} \left(x^2 f'^2+y^2 g'^2+2\right) & x f' & y g' & \frac{1}{2} \left(-x^2 f'^2-y^2 g'^2\right) \\
 x f' & 1 & 0 & -x f' \\
 y g' & 0 & 1 & -y g' \\
 \frac{1}{2} \left(-x^2 f'^2-y^2 g'^2\right) & -x f' & -y g' & \frac{1}{2} \left(x^2 f'^2+y^2 g'^2-2\right) \\
\end{array}
\right).
\end{equation}
Thus, if one preserves zero ISC (\ref{zeroiscgauge}) with the tetrad (\ref{TetWaveObintxyz}) one obtains the pair presenting the new gauge which differs from the gauge presented by the tetrad (\ref{tetwave}) and zero ISC (\ref{zeroiscgauge}). We call it as the Obukhov-Pereira-Rubilar gauge.

For the new gauge, by (\ref{K_A_A}) and (\ref{super_K}),  we get the teleparallel superpotential which in all coordinate indexes has non-zero components:
\begin{equation}
    \begin{array}{cccc}
        \sS{}_{0} {}^{0 1} =

\sS{}_{0} {}^{3 1} =

\sS{}_{3} {}^{1 0} =

\sS{}_{3} {}^{1 3} =
-\sS{}_{0} {}^{1 0} =

-\sS{}_{0} {}^{1 3} =

-\sS{}_{3} {}^{0 1} =

-\sS{}_{3} {}^{3 1} =
\frac{x f''}{f};

\\

\sS{}_{0} {}^{0 2} =

\sS{}_{0} {}^{3 2} =

\sS{}_{3} {}^{2 0} =

\sS{}_{3} {}^{2 3} =
-\sS{}_{0} {}^{2 0} =

-\sS{}_{0} {}^{2 3} =

-\sS{}_{3} {}^{0 2} =

-\sS{}_{3} {}^{3 2} =
\frac{y g''}{g}.
    \end{array}
\end{equation}
Taking the freely falling observer's proper vector (\ref{proper_FLRW}) in $(t,x,y,z)$ coordinates we get the Noether superpotential (\ref{super}):
\begin{equation}
    \begin{array}{cccc}
       {\scJ}{}^{0 1} ={\scJ}{}^{3 1} =-{\scJ}{}^{1 0}=-{\scJ}{}^{1 3}= \frac{x g f''}{8 \pi };

\\
{\scJ}{}^{0 2} = {\scJ}{}^{3 2} =-{\scJ}{}^{2 0}=-{\scJ}{}^{2 3}= \frac{y f g''}{8 \pi }.
    \end{array}
\end{equation}
Taking the divergence of the Noether superpotential we get the Noether current (\ref{curreqdivsupTEGR}):
\begin{equation}
  {\scJ}{}^{\mu} =   \left\{\frac{g f''+f g''}{8 \pi },0,0,\frac{g f''+f g''}{8 \pi }\right\}.
\end{equation}
It is zero due to the Einstein equation (\ref{einfg}).

\subsection{Zero current  in STEGR}

 Switching off gravity for the coordinates in (\ref{metwave}) in STEGR gives a gauge with the metric (\ref{metwave}) and zero STC (\ref{zeroconngauge}). This gives nonzero current (\ref{noethercurrdiag1000ST}), or (\ref{noethercurrdiagarbST}).  Here, we switch off gravity for the metric (\ref{MetWaveOb}) in the coordinates $(T,X,Y,Z)$.  It particularly gives $H(U,X,Y) = 0$ and, correspondingly, zero STC.
The transformed by (\ref{TXYZtotxyz}) metric (\ref{MetWaveOb}) goes to (\ref{metwave}), and zero STC is transformed as well (\ref{affconntransf})  and should be calculated as
\begin{equation}
\Gamma {}^\alpha {}_{\mu\nu} =   \frac{\partial x^\alf}{\partial X^\lambda} \frac{\partial}{\partial x^\mu} \l(\frac{\partial X^\lambda}{\partial x^\nu} \r),
\end{equation}
where $x^\mu \equiv (t,x,y,z)$, $X^\mu \equiv (T,X,Y,Z)$.
The non-zero components of the transformed STC are
\begin{equation}\label{connstegrOb}
   \begin{array}{cccc}
     \Gamma{}^{0} {}_{0 0} =
\Gamma{}^{0} {}_{3 3} =
\Gamma{}^{3} {}_{0 0} =
\Gamma{}^{3} {}_{3 3} =
-\Gamma{}^{0} {}_{0 3} =
-\Gamma{}^{0} {}_{3 0} =
-\Gamma{}^{3} {}_{0 3} =
-\Gamma{}^{3} {}_{3 0} =
\frac{1}{2} \left(x^2 f f'''+x^2 f' f''+y^2 g g'''+y^2 g' g''\right);
\\
\Gamma{}^{0} {}_{0 1} =
\Gamma{}^{0} {}_{1 0} =
\Gamma{}^{3} {}_{0 1} =
\Gamma{}^{3} {}_{1 0} =
-\Gamma{}^{0} {}_{1 3} =
-\Gamma{}^{0} {}_{3 1} =
-\Gamma{}^{3} {}_{1 3} =
-\Gamma{}^{3} {}_{3 1} =
x f f'';
\\
\Gamma{}^{0} {}_{0 2} =
\Gamma{}^{0} {}_{2 0} =
\Gamma{}^{3} {}_{0 2} =
\Gamma{}^{3} {}_{2 0} =
-\Gamma{}^{0} {}_{2 3} =
-\Gamma{}^{0} {}_{3 2} =
-\Gamma{}^{3} {}_{2 3} =
-\Gamma{}^{3} {}_{3 2} =
y g g'';
\\
\Gamma{}^{0} {}_{1 1} =
\Gamma{}^{3} {}_{1 1} =
f f';
\\
\Gamma{}^{0} {}_{2 2} =
\Gamma{}^{3} {}_{2 2} =
g g';
\\
\Gamma{}^{1} {}_{0 0} =
\Gamma{}^{1} {}_{3 3} =
-\Gamma{}^{1} {}_{0 3} =
-\Gamma{}^{1} {}_{3 0} =
\frac{x f''}{f};
\\
\Gamma{}^{1} {}_{0 1} =
\Gamma{}^{1} {}_{1 0} =
-\Gamma{}^{1} {}_{1 3} =
-\Gamma{}^{1} {}_{3 1} =
\frac{f'}{f};
\\
\Gamma{}^{2} {}_{0 0} =
\Gamma{}^{2} {}_{3 3} =
-\Gamma{}^{2} {}_{0 3} =
-\Gamma{}^{2} {}_{3 0} =
\frac{y g''}{g};
\\
\Gamma{}^{2} {}_{0 2} =
\Gamma{}^{2} {}_{2 0} =
-\Gamma{}^{2} {}_{2 3} =
-\Gamma{}^{2} {}_{3 2} =
\frac{g'}{g}.
\end{array}
\end{equation}
Thus, the new STEGR gauge is presented by the pair of the coordinates in (\ref{metwave}) with the connection (\ref{connstegrOb}). Using (\ref{connstegrOb}) we get the non-metricity (\ref{defQ}):
\begin{equation}
    \begin{array}{cccc}
     Q{}_{0 0 0} =

Q{}_{0 3 3} =

Q{}_{3 0 3} =

Q{}_{3 3 0} =
-Q{}_{0 0 3} =

-Q{}_{0 3 0} =

-Q{}_{3 0 0} =

-Q{}_{3 3 3} =

x^2 f f'''+x^2 f' f''+y^2 g g'''+y^2 g' g'';

\\

Q{}_{1 0 0} =

Q{}_{1 3 3} =

-Q{}_{1 0 3} =

-Q{}_{1 3 0} =
2 x f f'';

\\

Q{}_{2 0 0} =

Q{}_{2 3 3} =

-Q{}_{2 0 3} =

-Q{}_{2 3 0} =

2 y g g''.
    \end{array}
\end{equation}

Taking the freely falling observer's proper vector (\ref{proper_FLRW}) in $(t,x,y,z)$ coordinates we get: zero Komar superpotential (\ref{Komarsup}) which was found in previous sections; by  (\ref{addsupstegr}) one has zero additional (following from the divergence) part of Noether superpotential.
As a result, one has {zero total Noether superpotential in STEGR},
what gives zero Noether current.

\section{Concluding remarks}

Solutions for gravitational waves are quite important due to the modern epoch of detecting gravitational waves that has begun recently \cite{GW150914,GW151226,GW-A-LIGO}. For the best of our knowledge the energy characteristics of gravitational waves both in TEGR and in STEGR had not been studied enough up to now. In this article, using the developed by us earlier fully covariant formalism \cite{EKPT_2021,EPT:2022uij,EPT19,EPT_2020,EKPT_2021a} in TEGR and in STEGR, we have studied this problem on the example of a flat fronted exact (strong) gravitational wave of the only one polarization ``+’’.

The crucial property of our formalism both in TEGR and in STEGR is that a displacement vector $\xi^\alpha$ is included. Namely, interpretation of conserved quantities is defined by $\xi^\alpha$ that can be chosen as a Killing vector of spacetime, a proper vector of an observer, etc. Another important notion in the formalism \cite{EKPT_2021,EPT:2022uij,EPT19,EPT_2020,EKPT_2021a}  is a gauge, that is the equivalence class of pairs (tetrad, ISC) in TEGR, or pairs (coordinates, STC) in STEGR. To construct a physically sensible conserved quantity one has to find a corresponding gauge \cite{EKPT_2021,EPT:2022uij,EKPT_2021a}.

Since the gravitational wave solution is not stationary there are no timelike Killing vectors, formally it  means that it is impossible to construct energy density or energy density flux. Moreover as for the model under consideration we assume that gravitational wave fills infinite space (there is no static distant observers) it is impossible to construct conserved charges. However proper vectors of freely falling observers are determined easily and naturally. In correspondence with the equivalence principle, such observers have to measure zero. On the language of our formalism this means that components of the related current have to be zero. How we know, such a correspondence has not been stated up to now in teleparallel theories. In sections 4 and 5 here, we have closed this gap both in TEGR and in STEGR. We have found gauges when energetic characteristics of flat gravitational wave measured by a freely falling observer are zero that is just in a correspondence with the equivalence principle. It is the novelty and it is the main result of the paper. One could set a goal to find all possible gauges (not only particular ones, like here), which are in correspondence with the equivalence principle, but it is a more complicated task than the task studied here. Possibly, it will be considered in future.

What is interesting is that the simplest gauge used gives non-zero results, which in the limit of a weak wave coincides with the related pseudotensor formulae.
 Analogous results obtained in the non-covariant formalism are discussed in \cite{Formiga:2018maj,Formiga_2020}. We should stress here that  despite free falling masses can be used for detecting  non-zero energy of gravitational  waves, at least two masses separated by non-zero distance are needed. This means that formalism has to be adapted from the local one to a ``two-point'' formalism, or something analogous to this. Even for a simpler case of  Friedmann–Lemaître–Robertson–Walker cosmological metric consideration of distant masses makes energy issues
 more subtle, as it, for example, have been shown in the Harrison paper with a remarkable title ``Mining energy in an expanding Universe'' \cite{Harrison}. However, in our studies here we consider only one point-like moving geodesically in a gravitational wave metric, nevertheless, the obtained non-zero result has a physical sense at least for a weak wave. This means that the connection between the equivalence principle and the energy-momentum characteristics of a gravitational field studied here needs a deeper investigation.

 The present article can be developed in various directions. One can consider 1) a plane gravitational wave with two polarizations, 2) gravitational waves with cylindrical or spherical symmetry, 3) gravitational waves propagating inside matter. We leave such studies for a future work.

\appendix
\section{Arbitrary Lorentz rotations}
\setcounter{equation}{0}

Here, we formally derive the arbitrary Lorentz rotation dependent on $(t-z)$ only, which can be applied only to the tetrad as (\ref{lroth}) preserving ISC or only to the ICS as (\ref{spin_trans}) preserving tetrad. We present the compositions of simple Lorentz rotations dependent on $(t-z)$. It is claimed the note in subsection 4.1 that  there are no compositions such Lorentz rotations of the tetrad only preserving ISC (or of ISC only preserving tetrad) can change the Noether current.  %Then, connect the compositions of simple Lorentz rotations dependent on $(t-z)$ only with arbitrary Lorentz rotation dependent on $(t-z)$ only and conclude that if  no such compositions can change the Noether current, no such arbitrary Lorentz rotation can change the Noether current.

Arbitrary Lorentz rotation matrix  $\Lambda^a {}_b (t-z)$ of $SO(1,3)$ group can be expressed trough  $so(1,3)$ algebra
as
\begin{equation}\label{lamexp}
    \Lambda^a {}_b (t-z)=\exp (J_1 \alf_1 (t-z)+J_2 \alf_2 (t-z)+J_3 \alf_3 (t-z)+K_1 \beta_1 (t-z)+K_2 \beta_2 (t-z)+K_3 \beta_3 (t-z)),
\end{equation}
where $\alf_i (t-z),\beta_i (t-z)$ ($i=1,2,3$) are arbitrary functions and the generators of algebra $so (1,3)$ are
\begin{equation}
\begin{array}{cccc}
       J_1=i\l(\begin{array}{cccc}
      0  &  0  &  0  &  0   \\
     0  &  0  &  0  &  0   \\
     0  &  0  &  0  &  -1   \\
     0  &  0  &  1  &  0   \\
   \end{array} \r),   &     J_2=i\l(\begin{array}{cccc}
      0  &  0  &  0  &  0   \\
     0  &  0  &  0  &  1   \\
     0  &  0  &  0  &  0   \\
     0  &  -1  &  0  &  0   \\
   \end{array} \r), &    J_3=i\l(\begin{array}{cccc}
      0  &  0  &  0  &  0   \\
     0  &  0  &  -1  &  0   \\
     0  &  1  &  0  &  0   \\
     0  &  0  &  0  &  0   \\
   \end{array} \r),  \\
         K_1=i\l(\begin{array}{cccc}
      0  &  1  &  0  &  0   \\
    1  &  0  &  0  &  0   \\
     0  &  0  &  0  & 0   \\
     0  &  0  & 0  &  0   \\
   \end{array} \r),   &     K_2=i\l(\begin{array}{cccc}
      0  &  0  &  1  &  0   \\
     0  &  0  &  0  &  0   \\
     1  &  0  &  0  &  0   \\
     0  &  0  &  0  &  0   \\
   \end{array} \r), &    K_3=i\l(\begin{array}{cccc}
      0  &  0  &  0  &  1   \\
     0  &  0  &  0  &  0   \\
     0  &  0  &  0  &  0   \\
     1  &  0  &  0  &  0   \\
   \end{array} \r).  \\
\end{array}
\end{equation}
It is practically difficult to calculate directly the matrix exponent (\ref{lamexp}).
Therefore, we assume that there exist functions $\bar{\alf}_1 (t-z)$,  $\bar{\alf}_2 (t-z)$,  $\bar{\alf}_3 (t-z)$,  $\bar{\beta}_1 (t-z)$,  $\bar{\beta}_2 (t-z)$,  $\bar{\beta}_3 (t-z)$ such that
\begin{equation}\label{expexpexp}
    \exp (J_1 \alf_1 +J_2 \alf_2 +J_3 \alf_3 +K_1 \beta_1 +K_2 \beta_2 +K_3 \beta_3 ) = \exp (J_1 \bar{\alf}_1 ) \exp(J_2 \bar{\alf}_2 ) \exp(J_3 \bar{\alf}_3 ) \exp(K_1 \bar{\beta}_1 ) \exp(K_2 \bar{\beta}_2 ) \exp(K_3 \bar{\beta}_3 ),
\end{equation}
where  $\bar{\alf}_i \equiv \bar{\alf}_i (t-z),~\bar{\beta}_i \equiv \bar{\beta}_i (t-z),~{\alf}_i \equiv {\alf}_i (t-z),~{\beta}_i \equiv {\beta}_i (t-z)$, $i=1,2,3$. $\bar{\alf}_i,~\bar{\beta}_i$  can be connected with ${\alf}_i,~{\beta}_i$  using the Baker–Campbell–Hausdorff formula \cite{Jacobson1989}. This  formula gives direct expression for matrix $Z$ in $\exp (X) \exp (Y) = \exp (Z) $, as a function $Z=Z(X,Y)=\log (\exp X \exp Y)$. Applying this formula to each term under the exponent in (\ref{expexpexp}) sequentially, we can get the exact expressions of matrices $\bar{\alf}_i,\bar{\beta}_i$ as functions of matrices ${\alf}_i,{\beta}_i$. We don't need make these calculations directly here. It is enough for us to know that $\bar{\alf}_i,\bar{\beta}_i$ can be expressed explicitly through ${\alf}_i,{\beta}_i$ and then take the Lorentz rotation $ \Lambda^a {}_b (t-z)$ as a composition of simple Lorentz rotations:
\begin{equation}
\begin{array}{cccc}
 \Lambda^a {}_b (t-z)    = Exp (J_1 \bar{\alf}_1 ) \exp(J_2 \bar{\alf}_2 ) Exp(J_3 \bar{\alf}_3 ) Exp(K_1 \bar{\beta}_1 ) Exp(K_2 \bar{\beta}_2 ) Exp(K_3 \bar{\beta}_3 )=\\
   \Lambda_{(\bar{\alf}_1)} (t-z)   \Lambda_{(\bar{\alf}_2)} (t-z)  \Lambda_{(\bar{\alf}_3)} (t-z)\Lambda_{(\bar{\beta}_1)} (t-z)   \Lambda_{(\bar{\beta}_2)} (t-z)  \Lambda_{(\bar{\beta}_3)} (t-z)
\end{array}
\end{equation}
where $\bar{\alf}_i$, $\bar{\beta}_i$, $i=1,2,3$ are arbitrary functions of $(t-z)$ and
\begin{equation}
\begin{array}{cccc}
      \Lambda_{(\bar{\alf}_1)} (t-z)=  Exp(J_1 \bar{\alf}_1 (t-z) )=\l(\begin{array}{cccc}
      0  &  0  &  0  &  0   \\
     0  &  0  &  0  &  0   \\
     0  &  0  &  \cos [\bar{\alf}_1 (t-z)]  & -\sin [\bar{\alf}_1 (t-z)]    \\
     0  &  0  &  \sin [\bar{\alf}_1 (t-z)]   &  \cos [\bar{\alf}_1 (t-z)]    \\
   \end{array} \r)  , \\     \Lambda_{(\bar{\alf}_2)} (t-z)=  Exp(J_2 \bar{\alf}_2 (t-z) )=\l(\begin{array}{cccc}
      0  &  0  &  0  &  0   \\
     0  &  \cos [\bar{\alf}_2 (t-z)]  &  0  &  \sin [\bar{\alf}_2 (t-z)]   \\
     0  &  0  &  0  &  0   \\
     0  &  -\sin [\bar{\alf}_2 (t-z)]  &  0  &  \cos [\bar{\alf}_2 (t-z)]   \\
   \end{array} \r), \\     \Lambda_{(\bar{\alf}_3)} (t-z)=  Exp(J_3 \bar{\alf}_3 (t-z) )=\l(\begin{array}{cccc}
      0  &  0  &  0  &  0   \\
     0  &  \cos [\bar{\alf}_3 (t-z)]  &  -\sin [\bar{\alf}_3 (t-z)]  &  0   \\
     0  &  \sin [\bar{\alf}_3 (t-z)]  &  \cos [\bar{\alf}_3 (t-z)]  &  0   \\
     0  &  0  &  0  &  0   \\
   \end{array} \r),  \\
          \Lambda_{(\bar{\beta}_1)} (t-z)=  Exp(K_1 \bar{\beta}_1 (t-z) )=\l(\begin{array}{cccc}
       \cosh [\bar{\beta}_1 (t-z)]   &  \sinh [\bar{\beta}_1 (t-z)]   &  0  &  0   \\
    \sinh [\bar{\beta}_1 (t-z)]   &  \cosh [\bar{\beta}_1 (t-z)]   &  0  &  0   \\
     0  &  0  &  0  & 0   \\
     0  &  0  & 0  &  0   \\
   \end{array} \r) ,  \\         \Lambda_{(\bar{\beta}_2)} (t-z)=  Exp(K_2 \bar{\beta}_2 (t-z) )=\l(\begin{array}{cccc}
      \cosh [\bar{\beta}_2 (t-z)]   &  0  &  \sinh [\bar{\beta}_2 (t-z)]  &  0   \\
     0  &  0  &  0  &  0   \\
     \sinh [\bar{\beta}_2 (t-z)]   &  0  &  \cosh [\bar{\beta}_2 (t-z)]   &  0   \\
     0  &  0  &  0  &  0   \\
   \end{array} \r), \\       \Lambda_{(\bar{\beta}_3)} (t-z)=  Exp(K_3 \bar{\beta}_3 (t-z) )=\l(\begin{array}{cccc}
      \cosh [\bar{\beta}_3 (t-z)]   &  0  &  0  &  \sinh [\bar{\beta}_3 (t-z)]    \\
     0  &  0  &  0  &  0   \\
     0  &  0  &  0  &  0   \\
    \sinh [\bar{\beta}_3 (t-z)]   &  0  &  0  &  \cosh [\bar{\beta}_3 (t-z)]    \\
   \end{array} \r) . \\
\end{array}
\end{equation}
Arbitrary compositions of Lorentz rotations $  \Lambda_{(\bar{\alf}_i)} (t-z) $ and $  \Lambda_{(\bar{\beta}_i)} (t-z) $ ($i=1,2,3$) are used in calculations.

\section{Not freely falling tetrad}
\setcounter{equation}{0}

\setcounter{equation}{0}
Here, we show that the tetrad (\ref{TetWaveOb}) taken in \cite{Obukhov:2009gv} is not a freely falling tetrad and, thus, the observer that is rest in the frame  (\ref{TetWaveOb}) is not a freely falling.

The inversed tetrad (\ref{TetWaveOb}) is
\begin{equation}
  h_a{}^\mu   =  \left(
\begin{array}{cccc}
 -1+\frac{1}{2} H(T-Z,X,Y) & 0 & 0 & \frac{1}{2} H(T-Z,X,Y) \\
 0 & 1 & 0 & 0 \\
 0 & 0 & 1 & 0 \\
\frac{1}{2} H(T-Z,X,Y) & 0 & 0 & \frac{1}{2} H(T-Z,X,Y)+1 \\
\end{array}
\right).
\end{equation}
The time-like tetrad vector (which is taken as the observer's 4-velocity) is
\begin{equation}\label{Vel4Ob}
h_{\hat{0}} {}^{\mu}=\left\{-1+\frac{1}{2} H(T-Z,X,Y),0,0,\frac{1}{2} H(T-Z,X,Y)\right\},
\end{equation}
and, in \cite{Obukhov:2009gv}, it was assumed that (\ref{Vel4Ob}) is equal to the observer's 4-velocity, i.e.  the observer is rest in the frame  (\ref{TetWaveOb}).

Levi-Civita connection for (\ref{MetWaveOb}) is
\begin{equation}\label{ConnWaveObu}
    \begin{array}{cccc}
\cG{}^{0} {}_{0 0} =
\cG{}^{3} {}_{3 3} =
\cG{}^{3} {}_{0 0} =
\cG{}^{0} {}_{3 3} = \frac{1}{2} H^{(1,0,0)}(T-Z,X,Y);\\

\cG{}^{3} {}_{0 3} =
\cG{}^{3} {}_{3 0} =
\cG{}^{0} {}_{3 0} =
\cG{}^{0} {}_{0 3} = -\frac{1}{2} H^{(1,0,0)}(T-Z,X,Y);\\

\cG{}^{0} {}_{0 1} =
\cG{}^{1} {}_{0 0} =
\cG{}^{0} {}_{1 0} =
\cG{}^{1} {}_{3 3} =
\cG{}^{3} {}_{0 1} =
\cG{}^{3} {}_{1 0} = \frac{1}{2} H^{(0,1,0)}(T-Z,X,Y);\\

\cG{}^{0} {}_{1 3} =
\cG{}^{0} {}_{3 1} =
\cG{}^{1} {}_{0 3} =
\cG{}^{1} {}_{3 0} =
\cG{}^{3} {}_{1 3} =
\cG{}^{3} {}_{3 1} = -\frac{1}{2} H^{(0,1,0)}(T-Z,X,Y);\\

\cG{}^{0} {}_{0 2} =
\cG{}^{0} {}_{2 0} =
\cG{}^{2} {}_{0 0} =
\cG{}^{2} {}_{3 3} =
\cG{}^{3} {}_{0 2} =
\cG{}^{3} {}_{2 0} = \frac{1}{2} H^{(0,0,1)}(T-Z,X,Y);\\

\cG{}^{0} {}_{3 2} =
\cG{}^{0} {}_{2 3} =
\cG{}^{2} {}_{0 3} =
\cG{}^{2} {}_{3 0} =
\cG{}^{3} {}_{3 2} =
\cG{}^{3} {}_{2 3} = -\frac{1}{2} H^{(0,0,1)}(T-Z,X,Y),\\
    \end{array}
\end{equation}
where $U \equiv T-Z$, $H^{(1,0,0)}(U,X,Y)=\frac{\partial H(U,X,Y)}{\partial U}$, $H^{(0,1,0)}(U,X,Y)=\frac{\partial H(U,X,Y)}{\partial X}$, $H^{(0,0,1)}(U,X,Y)=\frac{\partial H(U,X,Y)}{\partial Y}$.
The geodesic equation has a form:
\begin{equation}\label{geodgen}
 l^\mu \equiv   \frac{d^2 x^\mu}{d \tau^2} + \cG {}^\mu {}_{\alpha \beta} \frac{dx^\alpha}{d\tau}\frac{dx^\beta}{d\tau} \equiv \frac{du^\mu}{d \tau} + \cG {}^\mu {}_{\alpha \beta} {u^\alpha}{u^\beta}\equiv \frac{\partial u^\mu}{\partial x^\kappa} u^\kappa + \cG {}^\mu {}_{\alpha \beta} {u^\alpha}{u^\beta} =0,
\end{equation}
where $u^\alpha$ is the observer's 4-velocity and the parameter $\tau$ can be taken as his proper time because he moves along the time-like curves.
Let's substitute (\ref{Vel4Ob}) identified with $u^\alpha$ and (\ref{ConnWaveObu}) into the left hand side of the equation (\ref{geodgen}).
If we  obtain zero,  the observer would be freely falling, but for general $H(U,X,Y)$ we obtain non zero result:
\begin{equation}\label{GeodWaveOb}
   l^\mu =  \left\{0,\frac{1}{2} H^{(0,1,0)}(U,X,Y),\frac{1}{2} H^{(0,0,1)}(U,X,Y),0\right\}.
\end{equation}
So, the observer  in \cite{Obukhov:2009gv} is not freely falling. Therefore, zero energetic characteristics of the wave obtained in \cite{Obukhov:2009gv}  with the use of (\ref{PereiraCurrdivsup}) (or (\ref{PereiraCurr})) cannot be interpreted as a correspondence with the equivalence principle.

{\bf Acknowledgments} AP has been supported by the Interdisciplinary Scientific and
Educational School of Moscow University “Fundamental and Applied Space Research”; EE
and AT are supported by RSF grant 21-12-00130. AT also thanks the Russian Government Program of Competitive Growth of Kazan Federal University.
The authors are grateful to Alexei Starobinsky  for the idea to consider gravitational wave solution in teleparallel gravity under the application of the Noether formalism.

\bibliography{references}
\bibliographystyle{Style}

\end{document}